\documentclass[superscriptaddress,dvipsnames,nofootinbib,amsmath,amssymb,aps,prd,floatfix,reprint]{revtex4-2}

\usepackage[titletoc,toc,title]{appendix}
\usepackage{mathtools}
\usepackage{amsmath, amsfonts,bm}
\usepackage[utf8]{inputenc}
\usepackage{xcolor}
\usepackage{graphicx}
\usepackage{float}
\usepackage{subfig}
\usepackage{comment}
\newcommand\myshade{85}
\colorlet{mylinkcolor}{RoyalPurple}
\colorlet{mycitecolor}{WildStrawberry}
\colorlet{myurlcolor}{BlueViolet}
\usepackage{orcidlink}

\usepackage[normalem]{ulem}

\usepackage{hyperref}
\usepackage[capitalise]{cleveref}

\hypersetup{
  linkcolor  = mylinkcolor!\myshade!black,
  citecolor  = mycitecolor!\myshade!black,
  urlcolor   = myurlcolor!\myshade!black,
  colorlinks = true,
}

\DeclareMathAlphabet{\mathup}{OT1}{\familydefault}{m}{n}

\newcommand{\be}{\begin{equation}} 
\newcommand{\ee}{\end{equation}}

\usepackage{array}
\newcommand{\PreserveBackslash}[1]{\let\temp=\\#1\let\\=\temp}
\newcolumntype{C}[1]{>{\PreserveBackslash\centering}p{#1}}
\newcolumntype{R}[1]{>{\PreserveBackslash\raggedleft}p{#1}}
\newcolumntype{L}[1]{>{\PreserveBackslash\raggedright}p{#1}}

\usepackage{multirow}
\usepackage{bm}


\crefname{equation}{Eq.}{Eqs.}
\crefname{section}{Section}{Sections}
\crefname{figure}{Fig.}{Figs.}
\crefname{table}{Table}{Tables}
\crefname{appendix}{Appendix}{Appendices}
\Crefname{figure}{Figure}{Figures}
\Crefname{equation}{Equation}{Equations}
\Crefname{section}{Section}{Sections}
\Crefname{table}{Table}{Tables}

\definecolor{llgray}{gray}{0.93}
\definecolor{lgray}{gray}{0.83}
\definecolor{deepmagenta}{rgb}{0.8, 0.0, 0.8}
\definecolor{ballblue}{rgb}{0.13, 0.67, 0.8}
\definecolor{celestialblue}{rgb}{0.29, 0.59, 0.82}
\definecolor{RedWine}{rgb}{0.743,0,0}
\definecolor{DarkGreen}{rgb}{0,0.6,0}

\definecolor{crimson}{RGB}{220,20,60}

\usepackage{fontawesome5}
\newcommand{\github}[1]{\href{#1}{\faGithub}
}

\begin{document}

\title{A sound horizon independent measurement of $H_0$ from BOSS, DESI and DES Y3}

\author{Zhiyu Lu\orcidlink{0009-0001-3701-6650}}
\email{zhiyulu@mail.ustc.edu.cn}
\affiliation{Department of Astronomy, School of Physical Sciences, University of Science and Technology of China, Hefei, Anhui 230026, China}
\affiliation{CAS Key Laboratory for Research in Galaxies
and Cosmology, School of Astronomy and Space Science, University of Science and Technology of China, Hefei, Anhui 230026, China}

\author{Th\'eo Simon\orcidlink{0000-0001-7858-6441}}
 \email{theo.simon@umontpellier.fr}
\affiliation{Laboratoire Univers et Particules de Montpellier (LUPM), Centre national de la recherche scientifique (CNRS) et Universit\'e de Montpellier, Place Eug\`ene Bataillon, 34095 Montpellier C\'edex 05, France}

\author{Vivian Poulin\orcidlink{0000-0002-9117-5257}}%
\affiliation{Laboratoire univers et particules de Montpellier (LUPM), Centre national de la recherche scientifique (CNRS) et Universit\'e de Montpellier, Place Eug\`ene Bataillon, 34095 Montpellier C\'edex 05, France}

\author{Yifu Cai\orcidlink{0000-0003-0706-8465}}%
\affiliation{Department of Astronomy, School of Physical Sciences, University of Science and Technology of China, Hefei, Anhui 230026, China}
\affiliation{CAS Key Laboratory for Research in Galaxies
and Cosmology, School of Astronomy and Space Science, University of Science and Technology of China, Hefei, Anhui 230026, China}


\begin{abstract}
We present a sound horizon independent measurement of the Hubble parameter using a multiprobe large-scale structure analysis {in the $\Lambda$CDM model}.
Removing the dependency on the sound horizon with a rescaling procedure at the matter power spectrum level, we analyse the BOSS full-shape power spectrum and bispectrum (for the first time) using the effective field theory of large-scale structure up to one loop.
We combine this analysis with the auto- and cross-angular power spectra from the DESI Legacy Imaging Survey DR9, the $3 \times 2$pt analysis from DES Y3 {within EFTofLSS framework}, and the CMB gravitational lensing power spectrum from \textit{Planck} PR3.
Our baseline analysis, which does not rely on supernovae data, yields $h = 0.702^{+0.022}_{-0.024}$, $\Omega_m = 0.310 \pm 0.013$, and $\sigma_8 = 0.799 \pm 0.020$, corresponding to $3-4 \%$ precision measurements.
When adding supernovae data from Pantheon+, we obtain a $2.6 \%$ measurement of $h$, with $h = 0.686 \pm 0.018$.
We further note that our EFTBOSS analysis indicates a slight deviation of the BAO scale parameter {$\alpha_{r_s}$ }(at $1.8 \sigma$) from its $\Lambda$CDM value, caused by the small scales of the bispectrum.
We finally use the sound horizon-free EFTBOSS analysis as a diagnosis for the presence of new physics, finding
that our results are consistent with the recent hints of evolving dark energy.
\end{abstract}

\maketitle

\section{\label{sec:intro}Introduction}

\begin{figure*}[!htbp]
  \centering
  \includegraphics[width=1\textwidth]{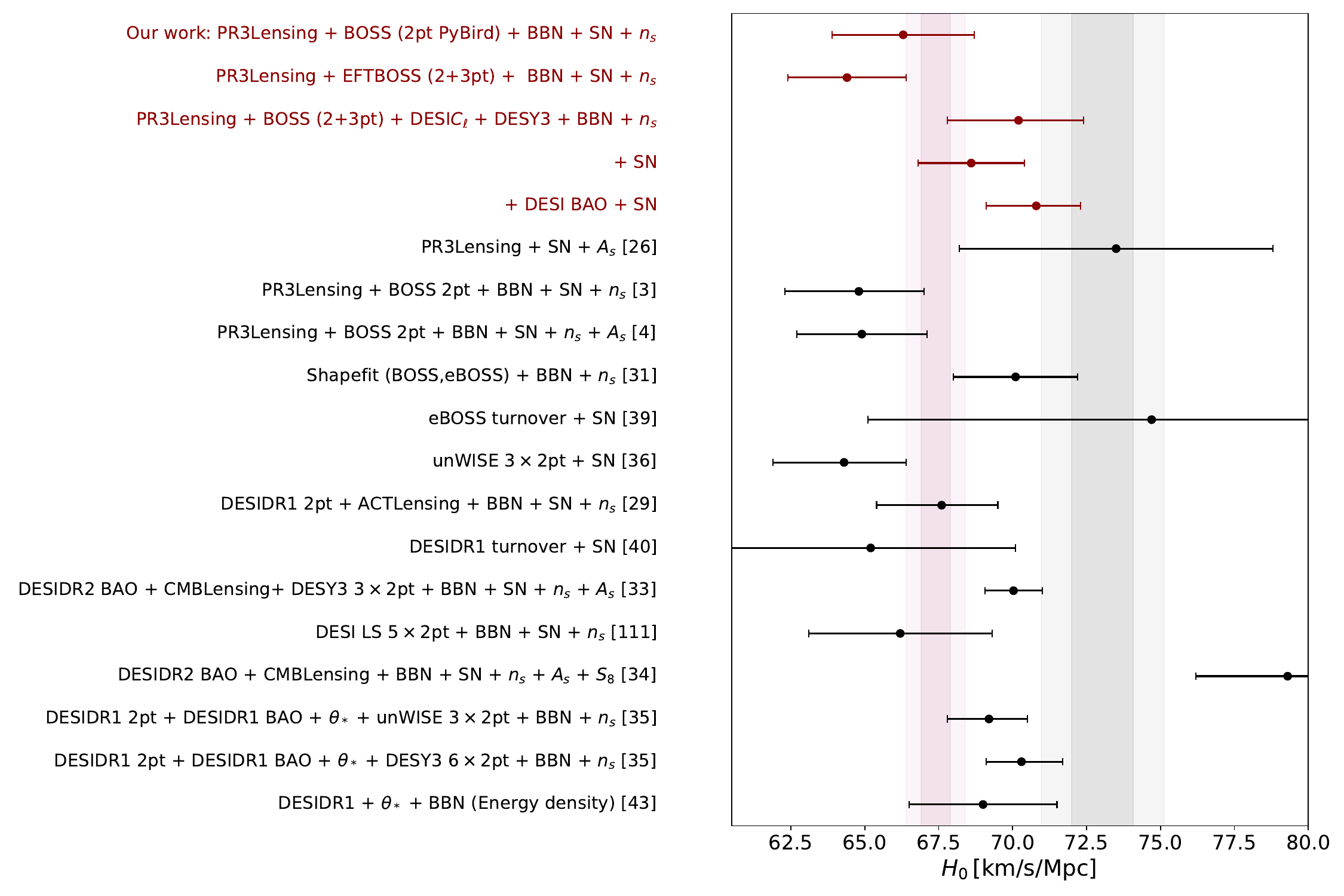}
  \caption{Summary of recent sound horizon-free determinations of the Hubble parameter $H_0$. In this plot, BAO data are always treated in a sound horizon-agnostic way, while ``$n_s$'', ``$A_s$'' and ``$S_8$'' indicate that Gaussian priors are imposed on these parameters (which are not necessarily the same for all analyses). The shaded bands denote the $1\sigma$ and $2\sigma$ regions from \textit{Planck}~\cite{Planck:2018vyg} (in pink) and SH0ES~\cite{Riess:2021jrx} (in grey). We note a good consistency with previous works~\cite{Philcox:2022sgj,Smith:2022iax} when using the same dataset, while noticeable differences appear when additional data such as the bispectrum, DESI$C_\ell$, SN, and DESI BAO are included.}
  \label{fig:h_ladder}
\end{figure*}

In recent years, several tensions between probes of the early and late universe have emerged~\cite{Cuesta:2014asa,BOSS:2014hhw,H0DN:2025lyy}.
Among these challenges, the \emph{Hubble tension} has attracted particular attention. This corresponds to a $\sim 5 \sigma$ tension between the early-time determination of this parameter from the \textit{Planck} collaboration, giving $H_0=67.36\pm 0.54\,{\rm km/s/Mpc}$~\cite{Planck:2018vyg}, and the local determination of this parameter from the SH0ES team~\cite{Riess:2021jrx}, using a (almost) cosmological model-independent calibration of supernova distances, yielding $H_0=73.5\pm0.81\,{\rm km/s/Mpc}$~\cite{H0DN:2025lyy}. 
At the same time, {several measurements relying on the sound horizon scale $r_s$--which encodes the acoustic physics prior to CMB--without involving CMB anisotropy data}, favor constraints on $H_0$ that are compatible with the CMB measurements, whether they come from \textit{Planck}~\cite{Planck:2018vyg}, ACT~\cite{AtacamaCosmologyTelescope:2025blo} or SPT~\cite{SPT-3G:2025bzu}.
This is for instance the case of the DESI DR2 BAO measurements, combined with a Big Bang Nucleosynthesis (BBN) prior on $\omega_b$, which gives $H_0=68.51\pm0.58\,{\rm km/s/Mpc}$~\cite{DESI:2025zgx}. 
The constraining power of these measurements is largely tied to the sound horizon $r_s$, motivating the development of sound horizon independent approaches.
Another intriguing cosmological conundrum, is a less significant but older $\sim 2-3 \sigma$ tension between the weak lensing~\cite{DES:2026fyc,Kilo-DegreeSurvey:2023gfr,KiDS:2020suj} and CMB determinations of the $S_8$ parameter~\cite{Planck:2018vyg,AtacamaCosmologyTelescope:2025blo}, which quantifies the amplitude of matter clustering (see Ref.~\cite{Pantos:2026koc} for a recent review). Even if recent reanalyses of the KiDS Legacy data~\cite{Wright:2025xka} report a significantly reduced tension, attributed to improved photometric redshift calibration and updated analysis pipelines, the DES Y6 collaboration still report a $2.6 \sigma$ tension~\cite{DES:2026fyc}.
This motivates further investigations using complementary large-scale structure (LSS) probes, even though current galaxy clustering determinations of this parameter, for instance from SDSS~\cite{Simon:2022csv,eBOSS:2020uxp} or DESI~\cite{DESI:2024hhd}, are compatible with both the weak lensing and CMB.
More recently, a $\sim 3 \sigma$ tension has emerged on the matter abundance today $\Omega_m$ between supernovae data (from Pantheon+~\cite{Brout:2022vxf}, Union3~\cite{Rubin:2023jdq}, or DES Y5~\cite{DES:2024jxu,DES:2025sig}) and clustering data from DESI~\cite{DESI:2024hhd,DESI:2025zgx} {(see Fig.~10 of Ref.~\cite{DESI:2025zgx})}.

In this work, we explore how a multiprobe LSS analysis can independently constrain these quantities, by exclusively considering LSS likelihoods that do not rely on the information from the sound horizon. {Instead, we anchor the inference of the Hubble constant to the matter–radiation equality scale. Although our datasets are not directly sensitive to the turnover in the matter power spectrum, the change in the growth rate of structure around matter–radiation equality leaves an imprint on the small-scale ($k > k_{\rm eq}$) clustering amplitude. This imprint can then be used to calibrate $H_0$~\cite{Smith:2022iax}.
}
This allows us to perform a non-trivial consistency test of the $\Lambda$CDM model by comparing sound horizon-based analyses with sound horizon-free analyses in order to isolate the impact of the sound horizon physics on the various cosmological discrepancies presented above.
In addition, such an analysis is a good diagnosis of the presence of physics beyond $\Lambda$CDM in the data. 
For instance, Refs.~\cite{Farren:2021grl,Philcox:2022sgj,Smith:2022iax} evaluate the internal consistency of SDSS/BOSS data~\cite{BOSS:2016wmc}, confronting sound horizon-based analyses with sound horizon-free analyses considering several types of new physics.
In particular, models which alter the sound horizon to address the Hubble tension~\cite{Poulin:2024ken} may not alter the $H_0$ determination from a sound horizon-free analysis (compared to $\Lambda$CDM). Thus, sound horizon-free methods provide a potential constraint on such models~\cite{Smith:2022iax}.


A number of works has already derived cosmological constraints that avoid explicit dependence on the sound horizon $r_s$.~\footnote{In this paper, we use $r_s$ to denote the sound horizon at the baryon drag epoch, defined as 
\begin{equation}
    r_s=\int_{z_d}^\infty\frac{c_s(z)}{H(z)}dz\,,
\end{equation}
where $z_d$ is the redshift at which photon-baryon decouples. 
}
For instance, Ref.~\cite{Baxter:2020qlr} demonstrated that CMB lensing provides an approximately $r_s$-independent probe, since this observable is integrated along the line of sight, significantly suppressing the signal from acoustic oscillations. Using this approach, they obtained $H_0 = 73.5 \pm 5.3\,{\rm km\,s^{-1}\,Mpc^{-1}}$ {from \textit{Planck} PR3}, with uncertainties large enough to be consistent with both early- and late-time measurements.
Subsequent studies extended this idea to galaxy surveys.
Ref.~\cite{Philcox:2020xbv} measured $H_0 = 65.6^{+3.4}_{-5.5}\,{\rm km\,s^{-1}\,Mpc^{-1}}$ using large-scale structure data from SDSS/BOSS by removing the prior on $\omega_b$, which otherwise constrains the sound horizon.
Further refinements consider the wiggle-no-wiggle split procedure of the matter power spectrum \cite{Farren:2021grl,Ghaemi:2025lgu,Zaborowski:2024car,Smith:2022iax}, which isolates the broadband shape from BAO features, improving the sound horizon-free constraint on $H_0$ from large-scale structure data (together with a BBN prior on $\omega_b$).
The EFTofLSS full modeling of the broadband shape yields $H_0 = 67.1^{+2.5}_{-2.9}\,{\rm km\,s^{-1}\,Mpc^{-1}}$~\cite{Philcox:2022sgj} from the BOSS galaxy power spectra.
Alternatively, the \texttt{Shapefit} method models the broadband shape of the galaxy power spectrum using compressed parameters, yielding $H_0 = 70.1 \pm 2.1\,{\rm km\,s^{-1}\,Mpc^{-1}}$ for BOSS + eBOSS \cite{Brieden:2021edu,Brieden:2022heh}, which is slightly higher than the EFTofLSS full modeling of the BOSS broadband shape.
More recently, Ref.~\cite{Zaborowski:2024car} performed a similar EFTofLSS-based analysis with the galaxy power spectra from DESI DR1, and found $H_0 = 67.9^{+1.9}_{-2.1}\,{\rm km\,s^{-1}\,Mpc^{-1}}$, compatible with the similar BOSS analysis, with a slight deviation (at $2 \sigma$) of the BAO scale parameter from its $\Lambda$CDM value when combined with \textit{Planck} lensing and Pantheon+, attributed to the discrepancy in $\Omega_m$ between Pantheon+ and  DESI.
Some other studies treat the sound horizon agnostically by taking $r_s$ as a free parameter~\cite{Pogosian:2020ded,GarciaEscudero:2025lef}, independent of $\omega_b$ and $\omega_m$, which normally set its value. The recent work~\cite{Sharma:2025iux} performed such an analysis using \textit{Planck}, DESI BAO, cosmic shear and supernovae data, finding a mild preference for dynamical dark energy and a large value of the Hubble rate parameter $H_0 = 74.7^{+3.4}_{-4.4}\,{\rm km\,s^{-1}\,Mpc^{-1}}$. 
{A similar approach is adopted in Ref.~\cite{Zaborowski:2025umc,Farren:2024rla}, which combines the sound horizon–independent EFTofLSS analysis of the DESI DR1 galaxy power spectra with the CMB acoustic scale $\theta_*$ from \textit{Planck} and the uncalibrated BAO measurements from DESI DR1 (which are agnostic to $r_s$).}
{Finally, let us note that other groups have attempted to directly measure (or forecast the precision of a measurement of) the matter-radiation equality scale to use it as a standard ruler~\cite{Cunnington:2022ryj,Lai:2025xxg,Bahr-Kalus:2023ebd,Bahr-Kalus:2025hhb}.  Moreover, another recent method attempt at directly {calculating} the total energy density (including the baryon-to-matter ratio) from large-scale structure data~\cite{Krolewski:2024jwj,Krolewski:2024epv,Krolewski:2025deb}.}
{We summarize in Fig.~\ref{fig:h_ladder} the recent sound horizon-free determinations of the Hubble parameter, while we refer to Ref.~\cite{Pantos:2026cxv} for a recent review on this topic. }

In this paper, we derive updated sound horizon-free cosmological constraints from large-scale structure data, using the power spectrum and bispectrum from BOSS, the $3 \times 2$pt analysis from DES Y3, the $C_\ell^{gg}$, $C_\ell^{\kappa g}$ and $C_\ell^{Tg}$ angular power spectra from the DESI Legacy Imaging Survey DR9, and the CMB gravitational lensing power spectrum from \textit{Planck} PR3. 
The paper is organised as follows.
In Sec.~\ref{sec_data_and_method}, we present the sound horizon-free likelihoods considered in this work, while in Sec.~\ref{sec:results} we present our main results. 
In particular, we first present the constraints from our individual likelihoods, we then discuss the cosmological implications of our combined analysis, before using our analysis as a diagnosis for beyond $\Lambda$CDM physics (considering early dark energy and dynamical dark energy cosmologies).
We finally conclude in Sec.~\ref{sec:conclusion}.

\section{Data and method} \label{sec_data_and_method}

\subsection{Datasets}\label{sec:data}

We perform cosmological parameter inference using Markov chain Monte Carlo (MCMC) sampling based on the Metropolis-Hastings algorithm implemented in \texttt{MontePython-v3}%
\footnote{\url{https://github.com/brinckmann/montepython_public}.}~\cite{Audren:2012wb,Brinckmann:2018cvx}, and 
interfaced with the Boltzmann solver \texttt{CLASS}%
\footnote{\url{https://github.com/lesgourg/class_public}.}~\cite{Blas:2011rf}.
The convergence of our MCMC chains is assessed using the Gelman-Rubin criterion $R-1<0.02$.
In our analysis, we consider the following likelihoods:

\begin{itemize}
    \item \textbf{Lensing:} The CMB-marginalized gravitational lensing spectrum from \textit{Planck} 2018~\cite{Planck:2018lbu}.

    \item \textbf{EFTBOSS:}
    The full-shape analysis of the power spectrum and bispectrum of BOSS luminous red galaxies (LRGs),
    modeled with the effective field theory of large-scale structure (EFTofLSS) up to one-loop order.\footnote{The first formulation of the EFTofLSS was carried out in Eulerian space in Refs.~\cite{Carrasco:2012cv,Baumann:2010tm} and in Lagrangian space in Ref.\cite{Porto:2013qua}. Once this theoretical framework was established, many efforts were made to improve this theory and make it predictive, such as the understanding of renormalization \cite{Pajer:2013jj, Abolhasani:2015mra}, the IR-resummation of the long displacement fields \cite{Senatore:2014vja, Baldauf:2015xfa, Senatore:2014via, Senatore:2017pbn, Lewandowski:2018ywf, Blas:2016sfa}, and the computation of the two-loop matter power spectrum \cite{Carrasco:2013sva, Carrasco:2013mua}. Then, this theory was developed in the framework of biased tracers (such as galaxies and quasars) in Refs. \cite{McDonald:2009dh, Senatore:2014eva, Mirbabayi:2014zca, Angulo:2015eqa, Fujita:2016dne, Perko:2016puo, Nadler:2017qto}, and extended to the bispectrum up to one-loop order in Refs.~\cite{DAmico:2022ukl,Philcox:2022frc}.}
    The SDSS-III BOSS DR12 galaxy sample is described in Ref.~\cite{BOSS:2016wmc}.
    The power spectrum and bispectrum measurements are taken from Ref.~\cite{DAmico:2022osl},
    where they were obtained using standard FKP-like estimators~\cite{Feldman:1993ky,Yamamoto:2005dz,Hand:2017irw,Bianchi:2015oia,1312.4611,Scoccimarro:2015bla,BOSS:2015npt,Gil-Marin:2016wya},
    as implemented in the \texttt{Rustico} code\footnote{\url{https://github.com/hectorgil/Rustico}.}~\cite{BOSS:2015npt,Gil-Marin:2016wya}.
    The covariance matrix, including cross-correlations between the power spectrum and bispectrum,
    is estimated from 2048 \texttt{Patchy} mock catalogs~\cite{Kitaura:2015uqa}, while the survey window function is measured using \texttt{fkpwin}\footnote{\url{https://github.com/pierrexyz/fkpwin}.}~\cite{Beutler:2018vpe}.
    The measurements are divided into two redshift bins, the CMASS (spanning a redshift range $0.2<z<0.43$ with $z_{\rm eff}=0.32$) and LOWZ (spanning a redshift range $0.43<z<0.7$ with $z_{\rm eff}=0.57$) catalogs%
    \footnote{\url{https://data.sdss.org/sas/dr12/boss/lss/}.}~\cite{BOSS:2015ewx}, that are further split into the north and south galactic skies.
    The analysis is restricted to mildly nonlinear scales, 
    with $k_{\rm max}=0.23\,h\,\mathrm{Mpc}^{-1}$ for CMASS and $k_{\rm max}=0.20\,h\,\mathrm{Mpc}^{-1}$ for LOWZ.
    We use an analytically marginalized likelihood~\cite{DAmico:2019fhj,DAmico:2020kxu}, adopting Jeffreys priors for the marginalized nuisance parameters to mitigate projection effects~\cite{Reeves:2025bxc,Hadzhiyska:2023wae}.In this paper, we use the \texttt{PyBird} likelihood~\cite{DAmico:2020kxu}, while in App.~\ref{sec:test_prior_and_code} we show the excellent agreement between this likelihood and \texttt{CLASS-PT}~\cite{Chudaykin:2020aoj,Philcox:2021kcw}, an alterntive EFT-based likelihood, in the framework of our sound horizon-free analysis.
    {In Fig.~\ref{fig:bestfit_vs_data}, we display the power spectrum and bispectrum predictions from \texttt{PyBird} of the BOSS CMASS NGC sample.}

    \item \textbf{DESI$C_\ell$:} 
    The auto- and cross- angular power spectra $C_{\ell}^{gg}$, $C_{\ell}^{\kappa g}$, and $C_{\ell}^{Tg}$ likelihood developped in Ref.~\cite{Lu:2025sjg}, where $g$ stands for the luminous red galaxies of the DESI Legacy Imaging Survey DR9~\cite{Zhou:2023gji,DESI:2022gle}, $\kappa$ stands for the \textit{Planck} PR4 Lensing map~\cite{Carron:2022eyg}, and $T$ stands for the \textit{Planck} PR4 temperature map~\cite{Planck:2020olo}.
    The data are divided into 4 photometric redshift bins ($z_{\rm eff} = \{ 0.470, \, 0.625, \, 0.785, \, 0.914 \}$)~\cite{Sailer:2024jrx,Kim:2024dmg}, measured by the \texttt{NaMaster} code\footnote{\url{https://github.com/LSSTDESC/NaMaster}.}~\cite{Alonso:2018jzx}. 
    The data are analyzed in the linear regime, in the multipole range $20\leq L \leq 243$, where we do not consider the Limber approximation as in Refs.~\cite{Sailer:2024jrx,Kim:2024dmg}, but use \texttt{Swift$C_\ell$}\footnote{\url{https://cosmo-gitlab.phys.ethz.ch/cosmo_public/swiftcl}.}~\cite{Reymond:2025ixl}, an accurate and differentiable JAX-based code computing the angular galaxy power spectrum beyond the Limber approximation using an FFTLog-based method.
    Our \texttt{MontePython-v3} likelihood, inspired from the \texttt{MaPar} likelihood\footnote{\url{https://github.com/NoahSailer/MaPar/tree/main}.} written in \texttt{Cobaya}~\cite{Sailer:2024jrx,Kim:2024dmg}, is publicly available here \github{https://github.com/GreenPlanck/DESICl}.
    We refer the reader to Ref.~\cite{Lu:2025sjg} for more details on our likelihood implementation, while we display in Fig.~\ref{fig:bestfit_vs_data} the galaxy-galaxy angular power spectra.

     \item \textbf{EFTDES:}  
     The EFTofLSS-based $3\times2$pt analysis of DES Y3 data from Ref.~\cite{DAmico:2025zui}, which combines galaxy-galaxy clustering, galaxy-galaxy lensing, and cosmic shear~\cite{DES:2016jjg}\footnote{\url{https://github.com/pierrexyz/pyfowl}}.
     After removing the correlations with the DESI angular power spectrum measurements (see below), the remaining information in the EFTDES likelihood is treated as statistically independent. {As done in Ref.~\cite{Zaborowski:2025umc} (see Sec.~3.5), we do not directly fit the DES data, but we consider Gaussian priors on the key cosmological parameters $\{ h, \, \Omega_m,\, \sigma_8 \}$ coming from the constraints of Ref.~\cite{DAmico:2025zui} (as shown in Tab.~\ref{tab:cosmo_results}) to (significantly) accelerate the convergence of our MCMC chains. }
     
     \item \textbf{External Priors:} BBN prior on the baryon density $\omega_b\sim\mathcal N(0.02268, 0.00038)$~\cite{Schoneberg:2019wmt},\footnote{This constraint is based on the theoretical prediction of Ref.~\cite{Consiglio:2017pot}, the experimental Deuterium fraction of Ref.~\cite{Cooke:2017cwo} and the experimental Helium fraction of Ref.~\cite{Aver:2015iza}.} and a \textit{Planck} prior on the primordial spectral index $n_s\sim \mathcal N(0.96,0.02)$~\cite{Planck:2018vyg}. In App.~\ref{sec:test_prior_and_code}, we instead consider the priors from Ref.~\cite{Zaborowski:2024car}, namely $\omega_b\sim\mathcal N(0.02218,0.00055)$ and $n_s\sim\mathcal N(0.9649,0.042)$, and show that it does not affect our conclusions.
\end{itemize}

Our baseline dataset consists of the combination of the above likelihoods.
We set uniform wide priors on the parameters $\{ \omega_{\rm cdm}, \, \ln 10^{10}A_s, \, h, \, \alpha_{r_s} \}$, corresponding to the physical dark matter density, the amplitude of the primordial power spectrum, the Hubble parameter $h \equiv H_0 / (100 \, {\rm km/s/Mpc})$, and the BAO location (see below). {Concerning the treatment of the neutrino masses, we consider two massless species and one species with $m_\nu = 0.06 e$V, following the convention of Ref.~\cite{Planck:2018vyg}.}

We note that in our analysis we neglect the correlation between EFTBOSS and DESI$C_\ell$ given that they
probe different cosmological volumes~\cite{Taylor:2022rgy,Maus:2025rvz,Ivanov:2026dvl}, as well as between EFTBOSS and EFTDES~\cite{Zaborowski:2025umc}.
However, approximately $20\%$ of the DESI Legacy Imaging Surveys footprint overlaps with the DES region used in the EFTDES likelihood~\cite{DESI:2018ymu,DES:2021wwk}. 
Since most of the constraining power of the DESI$C_\ell$ likelihood arises from the galaxy auto-spectrum $C_\ell^{gg}$, we remeasure this observable to remove the overlapping area with DES and recompute the covariance using~\texttt{NaMaster}~\cite{Alonso:2018jzx}. 
Given the low signal-to-noise ratio of $C_\ell^{\kappa g}$ {and $C_\ell^{T g}$} compared with $C_\ell^{g g}$, we further neglect the residual correlation between this observable and EFTDES, which is subdominant.
{Finally, when we combine DESI$C_\ell$ with the CMB lensing likelihood, we remove the overlaping region, namely $\ell < 243$, in the latter likelihood.}

\subsection{Method: a sound horizon-free analysis}

{In this subsection, we explain the extent to which our analysis is independent of information from the sound horizon.}
First, in order to remove the sound horizon information in the galaxy power spectrum and bispectrum of the EFTBOSS likelihood, we use the wiggle-no-wiggle split method~\cite{Farren:2021grl,Ghaemi:2025lgu,Zaborowski:2024car,Smith:2022iax}
\begin{equation}\label{eq:alpha_rs}
    P_{\rm lin}(k)=P_{\rm nw}(k)+P_{\rm w}(\alpha_{\rm r_s} k)\,,
\end{equation}
marginalizing over the parameter $\alpha_{r_s}$, where $P_{\rm nw}$ and $P_{\rm w}$ are the smooth and wiggle components of the power spectrum, respectively, determined by the cosmological parameters.
We extract the no-wiggle component {(\textit{i.e.}, the broadband shape)} by Fourier transforming the matter power spectrum into configuration space, 
then removing the BAO feature and interpolating the {broadband shape}, before transforming back to Fourier space,
following the procedure described in Refs.~\cite{Hamann:2010pw,Chudaykin:2020aoj}.
Although marginalizing over $\alpha_{r_s}$ removes the oscillatory BAO signal, residual sound horizon information can in principle enter through the baryon suppression scale, $\beta_{r_s}$.
Ref.~\cite{Farren:2021grl} shows
that, for current LSS surveys, the sound horizon information arising from the broadband shape is negligible.

Second, regarding the CMB gravitational lensing power spectrum, Ref.~\cite{Baxter:2020qlr} demonstrated that this observable is an approximately $r_s$-independent probe, since it is integrated along the line of sight, significantly suppressing the signal from acoustic oscillations.

Third, residual BAO information has also been shown to be negligible in the DESI$C_\ell$ likelihood.
In particular, the DESI$C_\ell$ analysis of Refs.~\cite{Reeves:2025axp, Reeves:2025xau} implements a wiggle–no-wiggle decomposition of the linear matter power spectrum similar to Eq.~\eqref{eq:alpha_rs}, before performing the projection into the galaxy angular power spectrum. They find that $\alpha_{r_s}$ remains only weakly constrained, confirming the effective sound horizon independence of this probe (see App.~B of Ref.~\cite{Reeves:2025xau}). 

Finally, regarding the EFTDES analysis~\cite{DAmico:2025zui}, it does not allow the BAO angular distance to be determined since (i) the BAO signal lies at larger angular separations, and (ii) the residual BAO signal is smoothed by the projection onto the line of sight (see App.~A of Ref.~\cite{Reeves:2025axp}).

\section{Cosmological results} \label{sec:results}

\begin{figure*}[!htbp]
  \centering
  \includegraphics[width=0.45\textwidth]{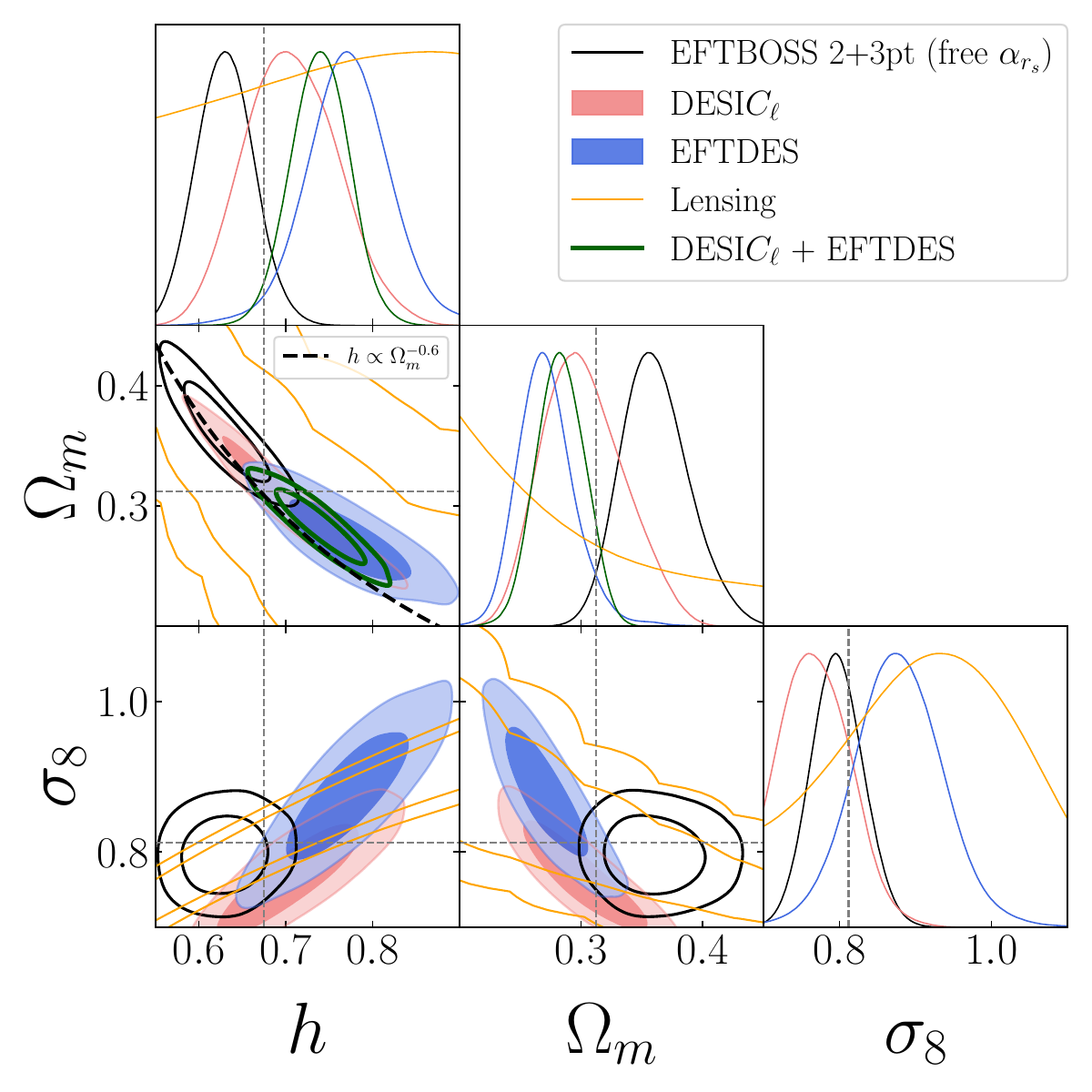}
  \includegraphics[width=0.45\textwidth]{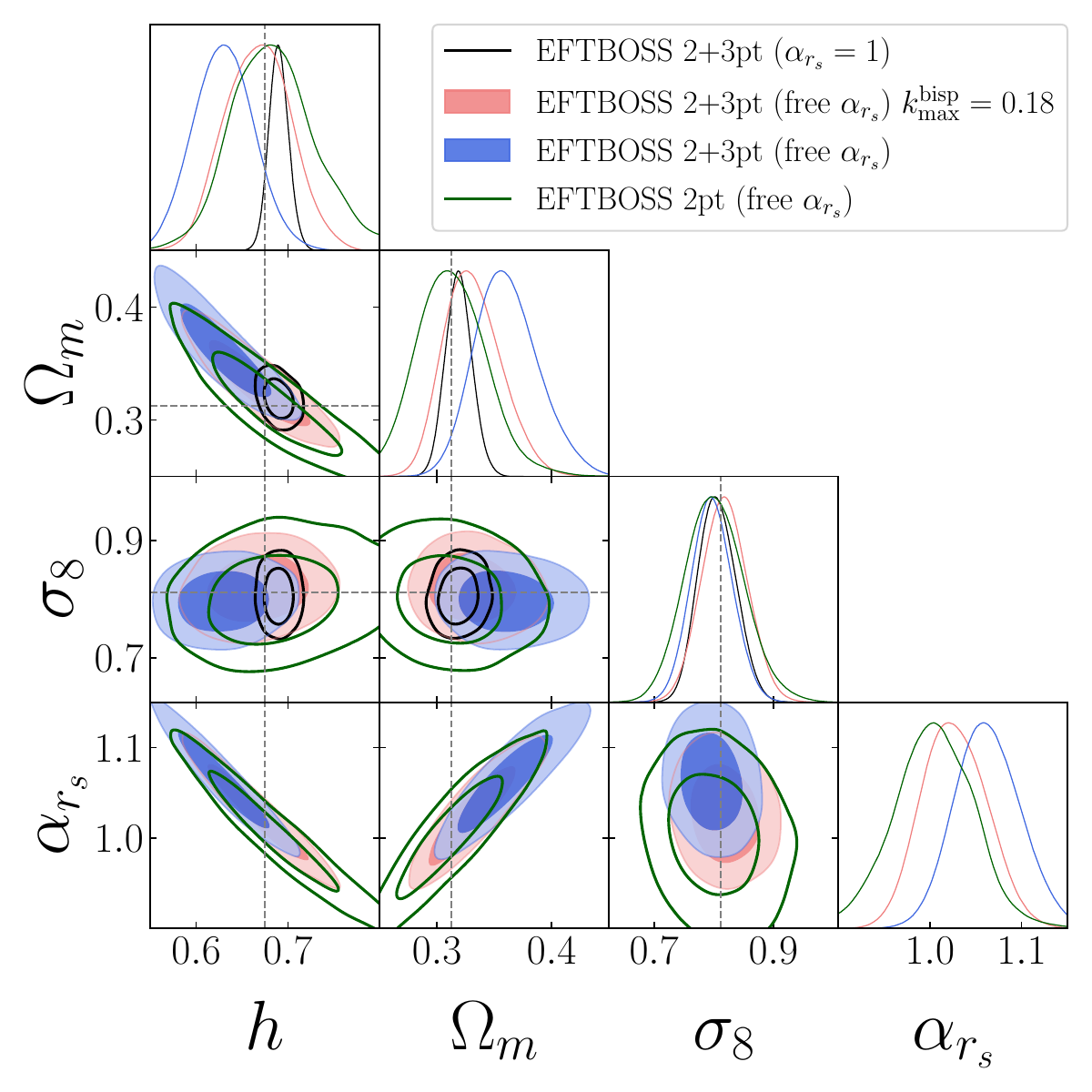}
  \includegraphics[width=0.45\textwidth]{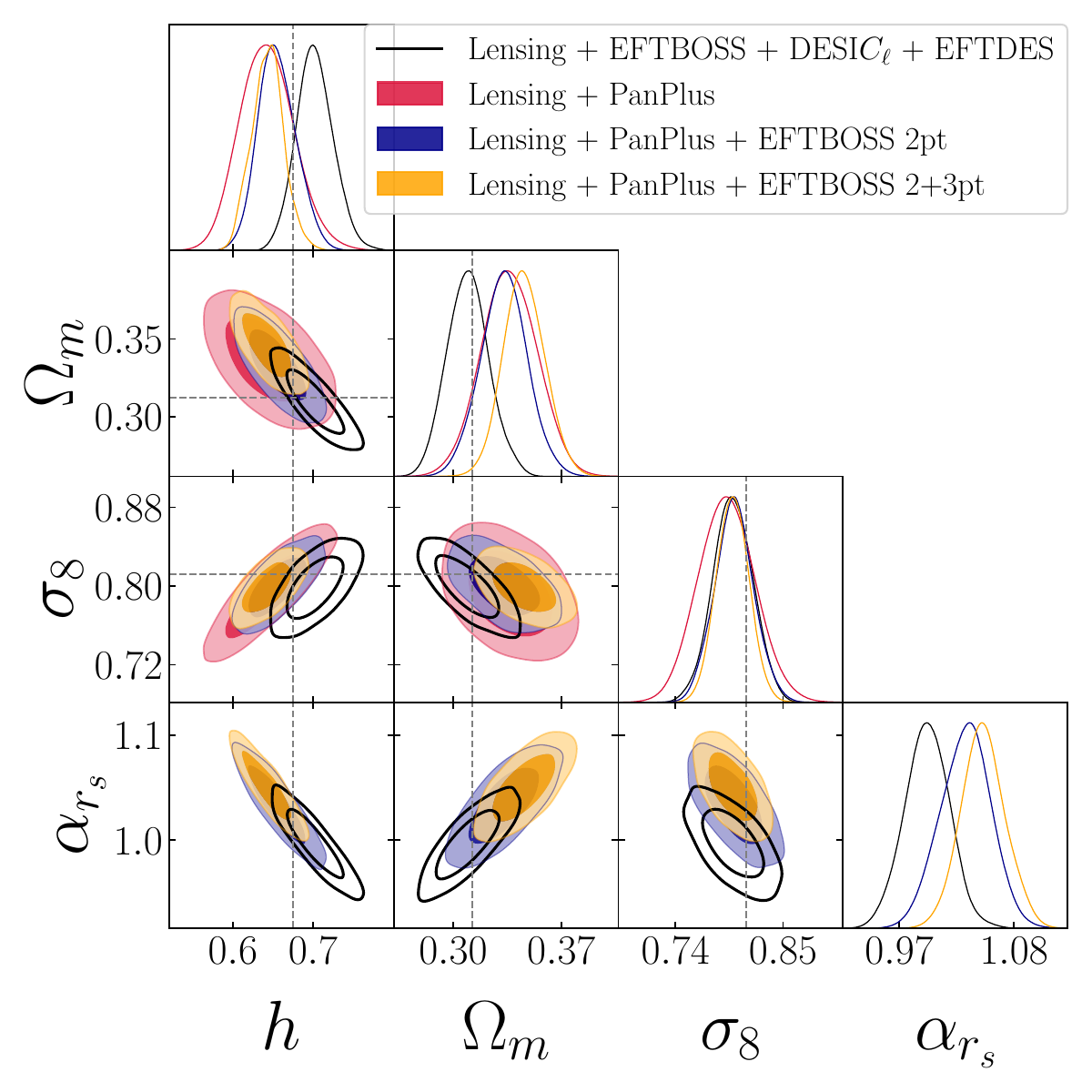}
  \includegraphics[width=0.45\linewidth]{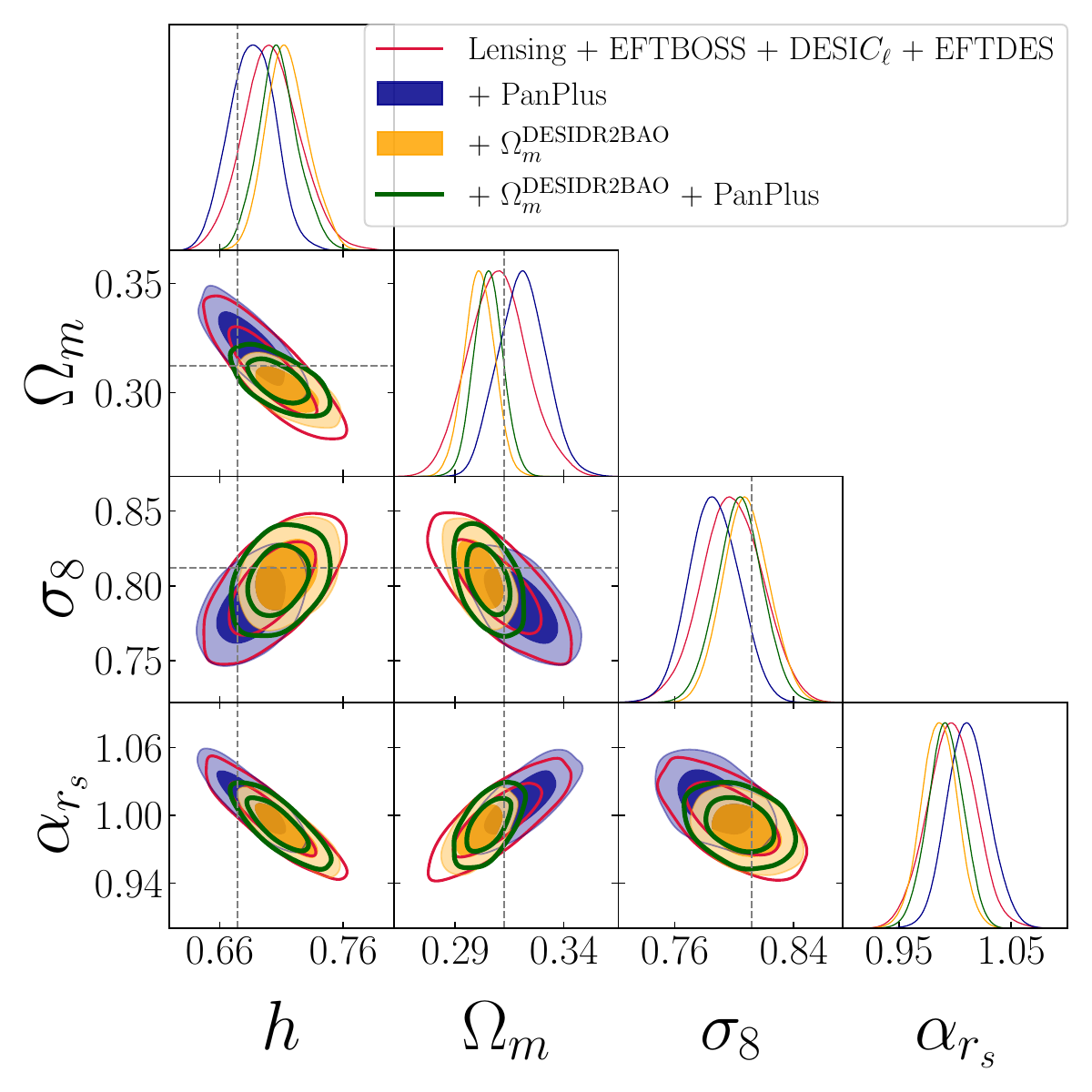}
  \caption{\textit{Top left} -- 2D posterior distributions from the individual sound horizon-free likelihoods considered in this work, with $\omega_b$ and $n_s$ fixed.
  \textit{Top right} -- 2D posterior distributions from several EFTBOSS configurations.
  \textit{Bottom left} -- 2D posterior distributions from several sound horizon-free likelihood combinations, using the prior on $n_s$ and $\omega_b$ defined in Sec.~\ref{sec:data}.
  \textit{Bottom right} -- 2D posterior distributions from several combinations between our baseline analysis, Pantheon+ and DESI DR2 BAO.
  In all panels, the dashed lines correspond to the \textit{Planck} mean values from TTTEEE + Lensing~\cite{Planck:2018vyg}.}
  \label{fig:mcmc_pieces_full}
\end{figure*}

\subsection{Internal consistency between the sound horizon-free likelihoods}\label{sec:internal_consistency}

\begin{table*}[!htpb]
    \centering
    \begin{tabular}{|c|c|c|c|}
        \hline
        & $h$ & $\Omega_m$ &  $\sigma_8$\\
        \hline
        DESI$C_\ell$ &  $0.706\pm 0.053$ & $0.302^{+0.029}_{-0.038}$ & $0.766\pm 0.048$ \\
        \hline
        EFTDES & $0.772\pm 0.049$  & $0.272^{+0.019}_{-0.026}$ &  $0.877\pm 0.059$\\
        \hline
        EFTBOSS & $0.630\pm 0.032$  & $0.361^{+0.025}_{-0.030}$ & $0.797\pm 0.034$ \\
        \hline
        EFTBOSS ($\alpha_{r_s}=1$) & $0.690\pm 0.011$&$ 0.319\pm 0.012$  & $ 0.806^{+0.030}_{-0.033}$\\
        \hline
        Lensing + DESI$C_\ell$ + EFTDES + EFTBOSS & $0.702^{+0.022}_{-0.024}$  & $0.310\pm 0.013$ & $0.799\pm 0.020$ \\
        \hline
        Lensing + DESI$C_\ell$ + EFTDES + EFTBOSS + PanPlus & $0.686\pm 0.018$  & $0.321\pm 0.011$ & $0.787\pm 0.017$ \\
        \hline
        Lensing + DESI$C_\ell$ + EFTDES + EFTBOSS + PanPlus + DESI BAO & $0.708^{+0.015}_{-0.017}$  & $0.3053\pm 0.0067$ & $0.804\pm 0.015$ \\
        \hline
        Lensing + PanPlus + EFTBOSS 2pt & $0.655^{+0.021}_{-0.026}$  & $0.333\pm 0.015$ & $0.800\pm 0.020$ \\
        \hline
        Lensing + PanPlus + EFTBOSS 2+3pt & $0.644\pm 0.020$  & $0.346\pm 0.014$ & $0.799\pm 0.017$ \\
        \hline
    \end{tabular}
    \caption{Cosmological results (posterior mean $\pm 68\%$ CL) from different combinations of sound horizon-free likelihoods. }
    \label{tab:cosmo_results}
\end{table*}

In the top left panel of Fig.~\ref{fig:mcmc_pieces_full}, we display the posterior distributions of $\{ h, \, \Omega_m, \, \sigma_8 \}$ obtained from each sound horizon-free likelihood introduced in Sec.~\ref{sec:data} --- namely, CMB lensing from \textit{Planck} 2018 (PR3), DESI$C_\ell$, EFTDES and EFTBOSS -- while in Tab.~\ref{tab:cosmo_results} we show the associated cosmological constraints.
For simplicity, we fix $n_s=0.965$~\cite{Planck:2018vyg} and $\omega_b=0.02235$~\cite{Schoneberg:2019wmt} in this subsection.

As shown in Fig.~\ref{fig:mcmc_pieces_full}, CMB lensing alone is not able to constrain $h$ and $\Omega_m$, but it forces the 2D posterior distribution of these parameters to follow the degeneracy direction $h \propto \Omega_m^{-0.6}$~\cite{Baxter:2020qlr}.
On the other hand, the DESI$C_\ell$, EFTDES and EFTBOSS likelihoods are able to set significant constraints on those parameters (determined at a $\lesssim 10 \%$ precision), while they all follow the CMB lensing degeneracy line in the $\{ h, \, \Omega_m \}$ plane.
However, those three likelihoods exhibit different directions in the $\{ \Omega_m, \, \sigma_8 \}$ and $\{ h, \, \sigma_8 \}$ planes, suggesting that a combination would break these degeneracies.
In addition, as shown in Tab.~\ref{tab:cosmo_results}, we note that EFTBOSS dominates the constraints on $h$ (by $\sim 70\%$ and $\sim 50 \%$ compared with DESI$C_\ell$ and EFTDES), as well as on $\sigma_8$ (by $\sim 40\%$ and $\sim 70\%$ compared with DESI$C_\ell$ and EFTDES), while EFTDES dominates the constraint on $\Omega_m$ by $\sim 40\%$ compared with DESI$C_\ell$ and EFTBOSS.
We further note that the DESI$C_\ell$ likelihood is consistent with both the EFTDES (at $1.7 \sigma$) and EFTBOSS (at $2.1 \sigma$) likelihoods, while the last two likelihoods are slightly inconsistent at $2.6 \sigma$, mostly coming from a shift along the $\{ h, \, \Omega_m \}$ degeneracy line.\footnote{Throughout this work, we quantify the tension between two analyses using the Gaussian tension metric~\cite{DESI:2025zgx,DAmico:2025zui} on the $\{ h, \, \Omega_m, \, \sigma_8 \}$ posterior  \begin{equation}
    T_N = \sqrt{(\mu_1 - \mu_2)^{\mathrm T} (\Sigma_1 + \Sigma_2)^{-1} (\mu_1 - \mu_2)} \, .
\end{equation}}
Finally, the EFTBOSS, EFTDES and DESI$C_\ell$ likelihoods are consistent with the \textit{Planck} (SH0ES~\cite{Riess:2021jrx}) reconstruction of $h$ at $1.4\sigma$ ($3.0\sigma$), $2.0\sigma$ ($0.5\sigma$) and $0.6\sigma$ ($0.8\sigma$), respectively.\footnote{We further note that the EFTBOSS, EFTDES and DESI$C_\ell$ likelihoods are consistent with \textit{Planck} (TTTEEE + Lensing)~\cite{Planck:2018vyg} at $ 1.8 \sigma$, $ 2.2 \sigma$ and $ 2.7 \sigma$ in the $\{h, \, \Omega_m, \, \sigma_8 \}$ posterior.}

\subsection{Internal consistency within the EFTBOSS likelihood}

To investigate the internal consistency of the sound horizon-free EFTBOSS likelihood, we first compare, in the top right panel of Fig.~\ref{fig:mcmc_pieces_full}, the analysis with and without the sound horizon marginalization (``free $\alpha_{r_s}$'' vs ``$\alpha_{r_s} =1$'').
The two analyses are slightly inconsistent, with a $1.8 \sigma$ shift in $h$ toward lower values and a $1.4 \sigma$ shift in $\Omega_m$ toward higher values.
Interestingly, these shifts are accompanied by a deviation of $\alpha_{r_s}$ from unity (or, in other words, from $\Lambda$CDM) at $1.8 \sigma$, with $\alpha_{r_s} = 1.063^{+0.034}_{-0.038}$.
As we discuss in Sec.~\ref{sec:new_physics}, this behavior may indicate some new physics.
This result differs from previous EFTofLSS sound horizon-free studies of BOSS data~\cite{Philcox:2022sgj,Smith:2022iax}, which include only the power spectrum, where no deviation was observed between the $\alpha_{r_s}$-free and $\alpha_{r_s}$-fix analyses (with $\alpha_{r_s} = 1.011^{+0.036}_{-0.028}$ in Ref.~\cite{Smith:2022iax} for example),\footnote{{However, we note that Ref.~\cite{Zaborowski:2024car} performed a similar analysis with the galaxy power spectra from DESI DR1, and found $q_{\rm BAO}<1$ (a parameter analogous to $\alpha_{r_s}$) at $2 \sigma$ when combined with \textit{Planck} lensing and Pantheon+. However, this deviation is driven by the discrepancy in $\Omega_m$ between Pantheon+ and  DESI.}}
indicating that the deviation found in our analysis comes from the bispectrum contribution.

Therefore, to quantify the impact of the bispectrum, we compare in the top right panel of Fig.~\ref{fig:mcmc_pieces_full} the EFTBOSS sound horizon-free analysis with the power spectrum only (``2pt''), with the one with both the power spectrum and the bispectrum (``2+3pt'').
In line with previous studies~\cite{Philcox:2022sgj,Smith:2022iax}, we report a constraint on $\alpha_{r_s}$ compatible with unity for the 2pt analysis, namely $\alpha_{r_s} = 1.002^{+0.049}_{-0.039}$.
Adding the bispectrum, we find an improvement of a factor of $\sim 2$ in the figure of merit (FoM)\footnote{Throughout the paper, in order to quantify an improvement in a 2D posterior distribution of two parameters, we use the ratio of the Figure of Merit (FoM)~\cite{Albrecht:2006um,DESI:2024hhd}, defined as FoM~$\propto |{\rm det} C|^{-1/2}$, where $C$ is the covariance matrix of the parameter posteriors.} of the $\{ h, \, \Omega_m \}$ plane.
The small shift between the $\alpha_{r_s}$-free and the $\alpha_{r_s}$-fix analyses is caused by the bispectrum contribution, as we observe a $1.0 \sigma$ shift in $h$ toward lower values and a $1.5 \sigma$ shift in $\Omega_m$ toward higher values when including the bispectrum.

\begin{figure*}[!htbp]
  \centering
  \includegraphics[width=0.45\textwidth]{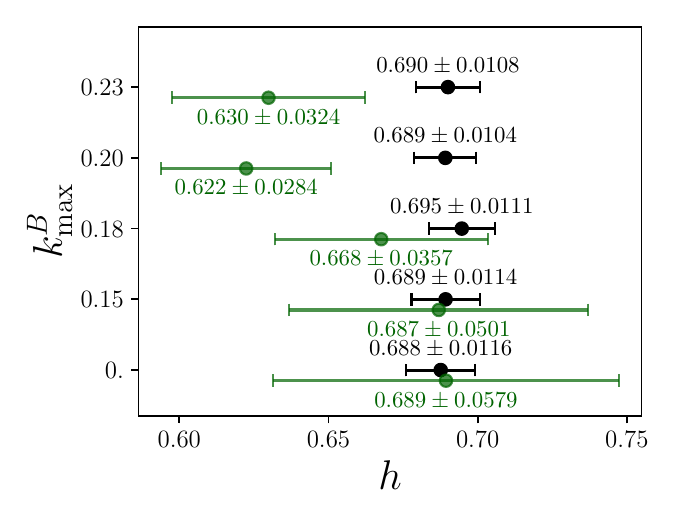}
  \includegraphics[width=0.45\textwidth]{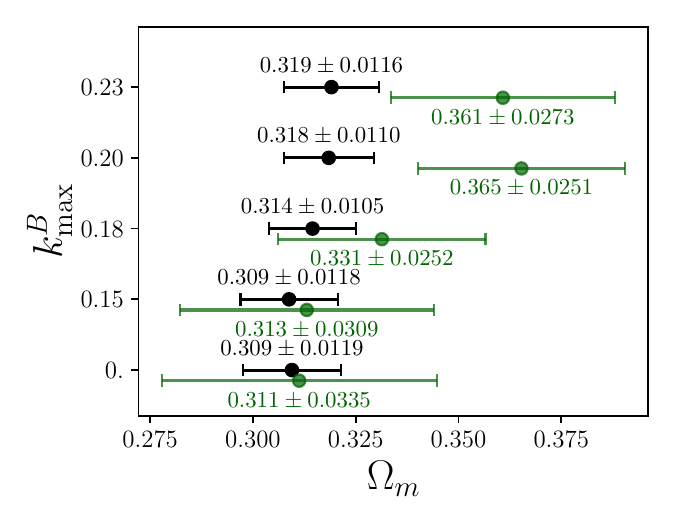}
  
  \includegraphics[width=0.45\textwidth]{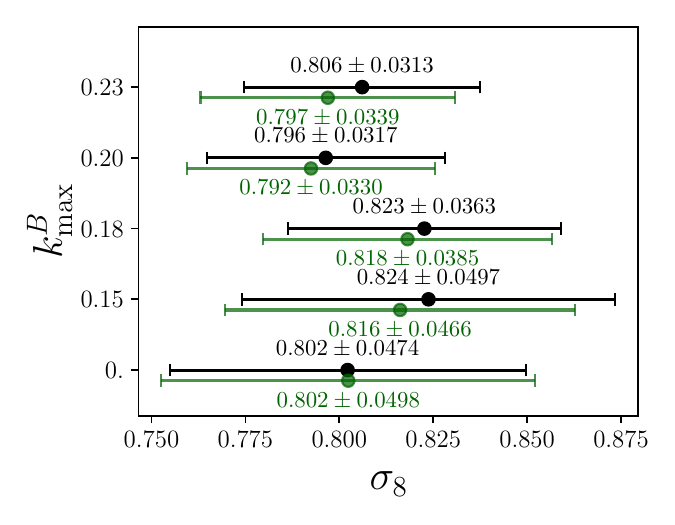}
  \includegraphics[width=0.45\textwidth]{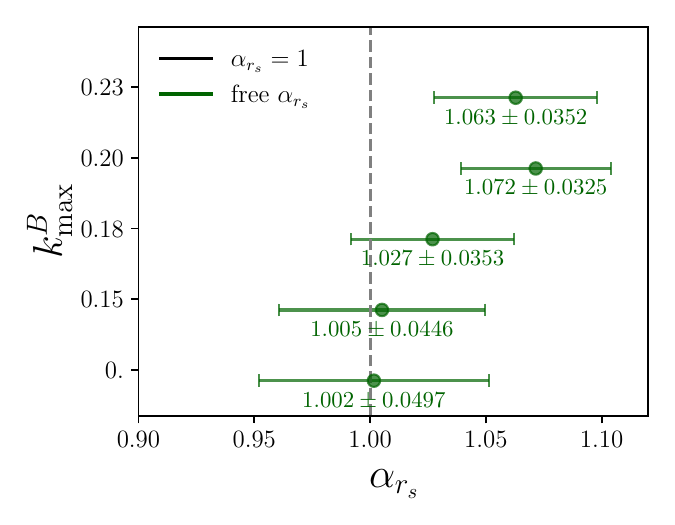}
  \caption{1D posterior distributions of $\{h, \, \Omega_m, \, \sigma_8, \, \alpha_{r_s} \}$ from the EFTBOSS 2+3pt analysis for several maximum bispectrum scales $k_{\rm max}^B$. We display the constraints of the analysis with (in green) and without (in black) the marginalization on the sound horizon.  }
  \label{fig:h_Om_varykmax}
\end{figure*}

To further investigate the impact of the bispectrum in the sound horizon-free EFTBOSS likelihood, we study in detail the impact of the choice of $k_{\rm max}^B$, the maximum bispectrum wavenumber, in both the $\alpha_{r_s}$-free and $\alpha_{r_s}$-fix analyses, as shown in Fig.~\ref{fig:h_Om_varykmax}.
While the constraints on $\{h, \, \Omega_m, \, \sigma_8  \}$ are quite stable for the $\alpha_{r_s}$-fix analysis, we observe a shift of $\sim 1.4 \sigma$ in $h $ and $\Omega_m$ between $k_{\rm max}^B = 0.18 \, h \, {\rm Mpc}^{-1}$ and $k_{\rm max}^B = 0.20 \, h\, {\rm Mpc}^{-1}$ for the $\alpha_{r_s}$-free analysis.\footnote{{We note that our bispectrum analysis includes 170 data points in total. Among them, 99 satisfy $k_1, k_2, k_3 < k_{\rm max}^B = 0.18 \, h{\rm Mpc}^{-1}$, while the remaining 71 configurations have at least one side with $k > k_{\rm max}^B = 0.18 \, h{\rm Mpc}^{-1}$.}}
These shifts are accompanied by an $\alpha_{r_s}$ constraint which becomes $\sim 2 \sigma$ away from unity. 
In addition, as shown in the top right panel of Fig.~\ref{fig:mcmc_pieces_full}, reducing the maximum bispectrum scale from $k_{\rm max}^B = 0.23\,h\,{\rm Mpc}^{-1}$ (corresponding to our baseline) to $k_{\rm max}^B = 0.18\,h\,{\rm Mpc}^{-1}$, mitigates the discrepancy between the $\alpha_{r_s}$-free and the $\alpha_{r_s}$-fix analyses from $1.8 \sigma$ to $0.6\sigma$ and restores consistency with $\alpha_{r_s} = 1$.
We further note that the inconsistency with EFTDES is reduced from $2.7 \sigma$ to $1.8\sigma$ with $k_{\rm max}^B = 0.18\,h\,{\rm Mpc}^{-1}$.

\subsection{Combination}

\begin{figure}
    \centering
    \includegraphics[width=1\linewidth]{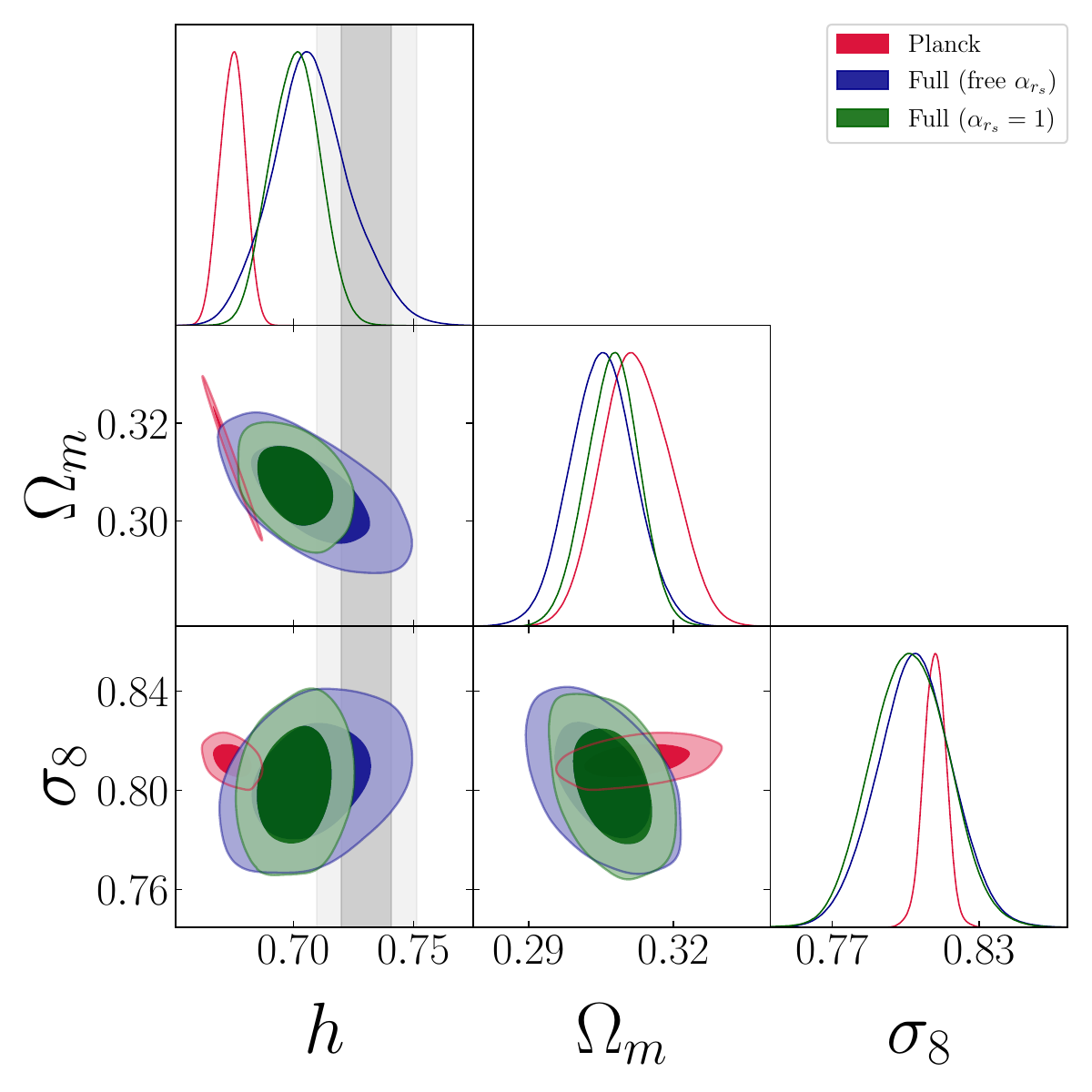}
    \caption{2D posterior distributions from the full dataset, namely Lensing + EFTBOSS (2+3pt) + DESI$C_\ell$ + EFTDES + PanPlus + $\Omega_m^{\rm DESIDR2BAO}$, for the $\alpha_{r_s}$-free and $\alpha_{r_s}$-fix analyses. For comparison, we display the posteriors from \textit{Planck}~\cite{Planck:2018vyg} in red and the SH0ES constraint on $H_0$~\cite{Riess:2021jrx} in grey shaded bands.}
    \label{fig:full_compare_marg_nomarg}
\end{figure}

The bottom left panel of Fig.~\ref{fig:mcmc_pieces_full} presents the constraints from different data combinations of the sound horizon-free likelihoods presented above.
Our baseline combination, namely Lensing + EFTBOSS + DESI$C_\ell$ + EFTDES, allows us to determine the three cosmological parameters $\{h, \, \Omega_m, \, \sigma_8 \}$ at a $3-4\%$ precision (see Tab.~\ref{tab:cosmo_results}). 
Our baseline combination is compatible with \textit{Planck} at $1.2 \sigma$, $0.2 \sigma$ and $0.6 \sigma$ for $h$, $\Omega_m$ and $\sigma_8$, respectively, while it is consistent with the SH0ES value of $h$~\cite{Riess:2021jrx} at $2.1 \sigma$ (see Fig.~\ref{fig:h_ladder}).
In addition, our results are compatible with $\Lambda$CDM, since we obtain a constraint on $\alpha_{r_s}$ compatible with unity, namely $\alpha_{r_s}= 0.997\pm 0.022$, and varying the choice of $k_{\rm max}^B$ has only a minor impact on the final reconstructed parameters.

For comparison, we also display, in the  bottom left panel of Fig.~\ref{fig:mcmc_pieces_full}, two previous sound horizon-free analyses, namely Lensing + PanPlus\footnote{We consider here the Pantheon+ likelihood of uncalibrated luminosity distance of type Ia supernovae (SNeIa) in the range $0.01 < z < 2.3$~\cite{Brout:2022vxf}. } (with a BBN prior on $\omega_b$ and a \textit{Planck} prior on $n_s$), as studied in  Ref.~\cite{Baxter:2020qlr}, as well as Lensing + PanPlus + EFBOSS 2pt (with free-$\alpha_{r_s}$), as performed in Refs.~\cite{Philcox:2022sgj,Smith:2022iax}.
Our baseline analysis improves the $\{ h, \, \Omega_m, \, \sigma_8$\} FoM by a factor of $\sim 3$ compared to the former analysis and by a factor of $\sim 1.5$ compared to the latter analysis.
In addition, to compare with previous $\alpha_{r_s}$-free BOSS studies, we show in Fig.~\ref{fig:mcmc_pieces_full} the impact of the EFTBOSS bispectrum information when added to  the Lensing + PanPlus + EFBOSS 2pt dataset, allowing us to improve the constraints by $15 \%$ on $h$ and $\Omega_m$.
We further note that while our baseline combination favors a relatively high value of $h$, the Lensing + PanPlus + EFTBOSS 2pt+3pt combination selects a fairly small value of $h = 0.643\pm 0.020$ because of the high $\Omega_m$ value preferred by Pantheon+. 
However, we note that across all combinations considered, we recover very consistent constraints on $\sigma_8 \sim 0.799\pm 0.020$.

The baseline constraints can be further improved by incorporating an external prior on $\Omega_m\sim\mathcal N(0.338,0.018)$ from Pantheon+ (as done in Ref.~\cite{Smith:2022iax}), as illustrated in the bottom right panel of Fig.~\ref{fig:mcmc_pieces_full}.
In particular, it tightens the constraints by a factor of $\sim 1.5$ (in the FoM of $\{h, \, \Omega_m, \, \sigma_8 \}$) while decreasing the value of $h$, giving $h = 0.686\pm 0.018$, reflecting the difference between the preferred matter density in supernova and large-scale structure data~\cite{Lu:2025sjg}.
Furthermore, Ref.~\cite{Zaborowski:2025umc} demonstrated that rescaling the BAO parameters $(\alpha_\parallel, \,  \alpha_\perp, \,  \alpha_{\rm iso})$ by $\alpha_{r_s}$ leaves the inferred value of $\Omega_m$ {from DESI BAO} unchanged, implying that this constraint arises entirely from the Alcock-Paczynski effect rather than the sound horizon.
Motivated by this result, we also apply a DESI DR2 BAO prior on $\Omega_m$ to our baseline dataset, namely $\Omega_m\sim\mathcal N(0.2975,0.0086)$, improving the constraints by a factor of $1.9$ and allowing a $2.4\%$ measurement on $h$ free from sound horizon and supernovae data (see the bottom right panel of Fig.~\ref{fig:mcmc_pieces_full}).
Finally, as shown in Fig.~\ref{fig:full_compare_marg_nomarg}, we combine the Pantheon+ and DESI DR2 BAO priors on $\Omega_m$ with our baseline dataset to obtain the tightest results of our work, namely
\begin{equation}
    \begin{aligned}
    &h = 0.708^{+0.015}_{-0.017}\,,\\
    &\Omega{}_{m } = 0.3053\pm 0.0067\,,\\
    &\sigma_8 = 0.804\pm 0.015\, ,
\end{aligned}
\end{equation}
corresponding to a $2.3 \%$, $2.2 \%$ and $1.9 \%$ precision measurements, respectively.
Those results are compatible with \textit{Planck} at $1.9 \sigma$, $0.7 \sigma$ and $0.5 \sigma$ for $h$, $\Omega_m$ and $\sigma_8$, and with SH0ES at $1.3 \sigma$ for $h$ (see Fig.~\ref{fig:h_ladder}).
{In Fig.~\ref{fig:full_compare_marg_nomarg}, we also display the constaints from our full dataset with $\alpha_{r_s}$-fix, yielding 
\begin{equation}
    \begin{aligned}
    &h = 0.701\pm 0.010\,,\\
    &\Omega{}_{m } = 0.3074\pm 0.0055\,,\\
    &\sigma_8 = 0.802\pm 0.015\, .
\end{aligned}
\end{equation}
The $\alpha_{r_s}$-free and $\alpha_{r_s}$-fix analyses are consistent up to $0.4 \sigma$ on the cosmological parameters, with $\alpha_{r_s}$ compatible with unity ($\alpha_{r_s} = 0.991\pm 0.016$) in the full analysis. }

\subsection{Looking for new physics}\label{sec:new_physics}

\begin{table*}[!htbp]
\centering
\begin{tabular}{|c|ccccccccccc|cc|}
\hline
&$\omega_b$ & $\omega_{\rm cdm}$ & $h$ & $\ln(10^{10}A_s)$ & $n_s$ & $\alpha_{r_s}$ & $f_{\rm axion}$ & $\theta_{\rm scf}$ & $\log_{10} a_c$ &$w_0$&$w_a$ &$\Omega_m$ & $\sigma_8$ \\
\hline
EDE&0.02253 & 0.1306 & 0.7219 & 3.0978 & 0.9889 & 1.0 & 0.122 & 2.83 & $-3.56$ & -1.0 & 0. & 0.3129 & 0.8126 \\
DDE& 0.02226 & 0.1187 & 0.6675 & 3.0442 & 0.9700 & 1.0 & - & - & - & -0.7779 & -0.7189 & 0.3177 & 0.8033  \\
\hline
\end{tabular}
\caption{\label{table:mock_cosmology} Fiducial cosmological parameters used to generate the mock data. The EDE cosmology is chosen following Ref.~\cite{Smith:2019ihp}, while the DDE cosmology is based on the bestfit of Ref.~\cite{Lu:2025sjg}.}
\end{table*}

\begin{figure*}[!htbp]
  \centering
  \includegraphics[width=0.45\textwidth]{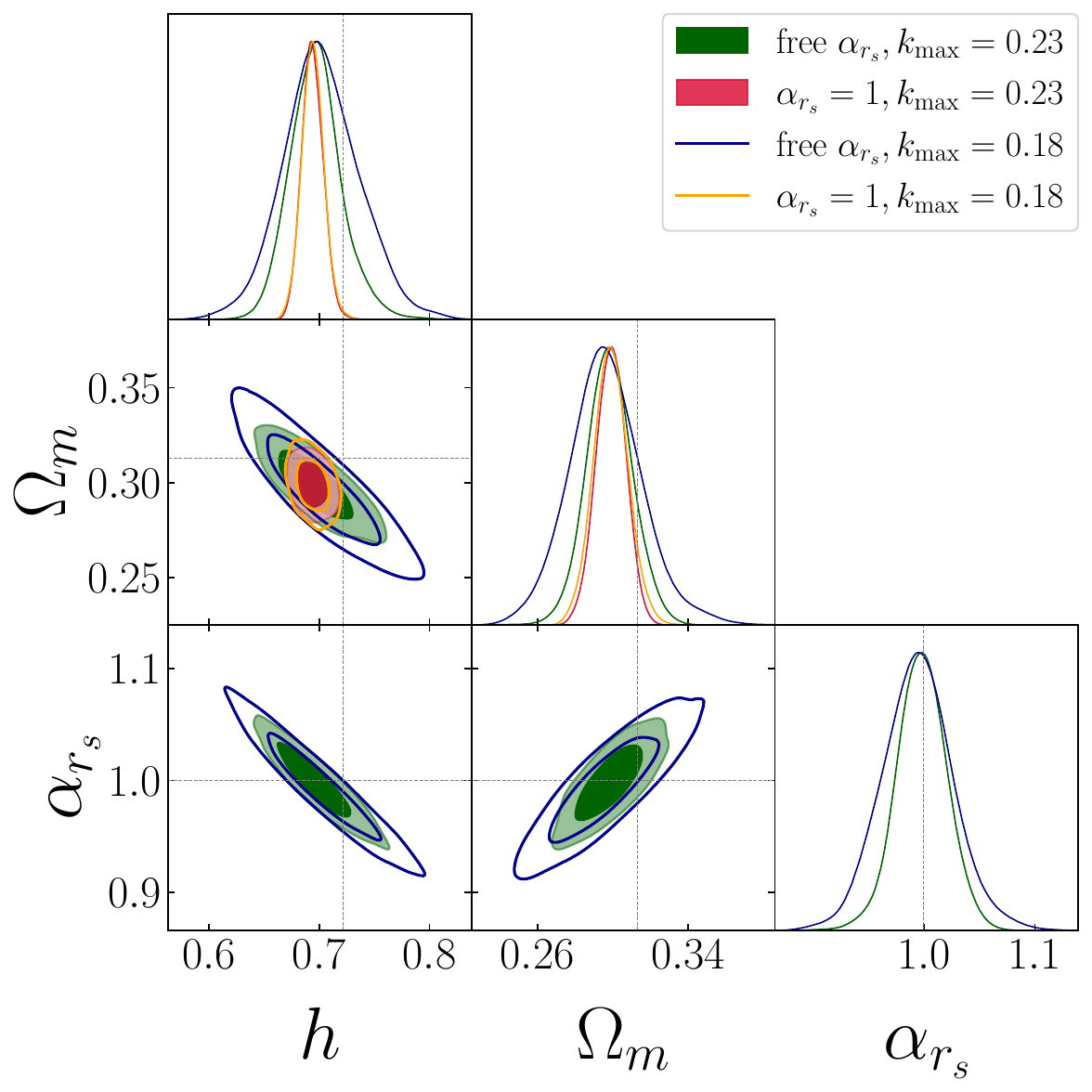}
  \includegraphics[width=0.45\linewidth]{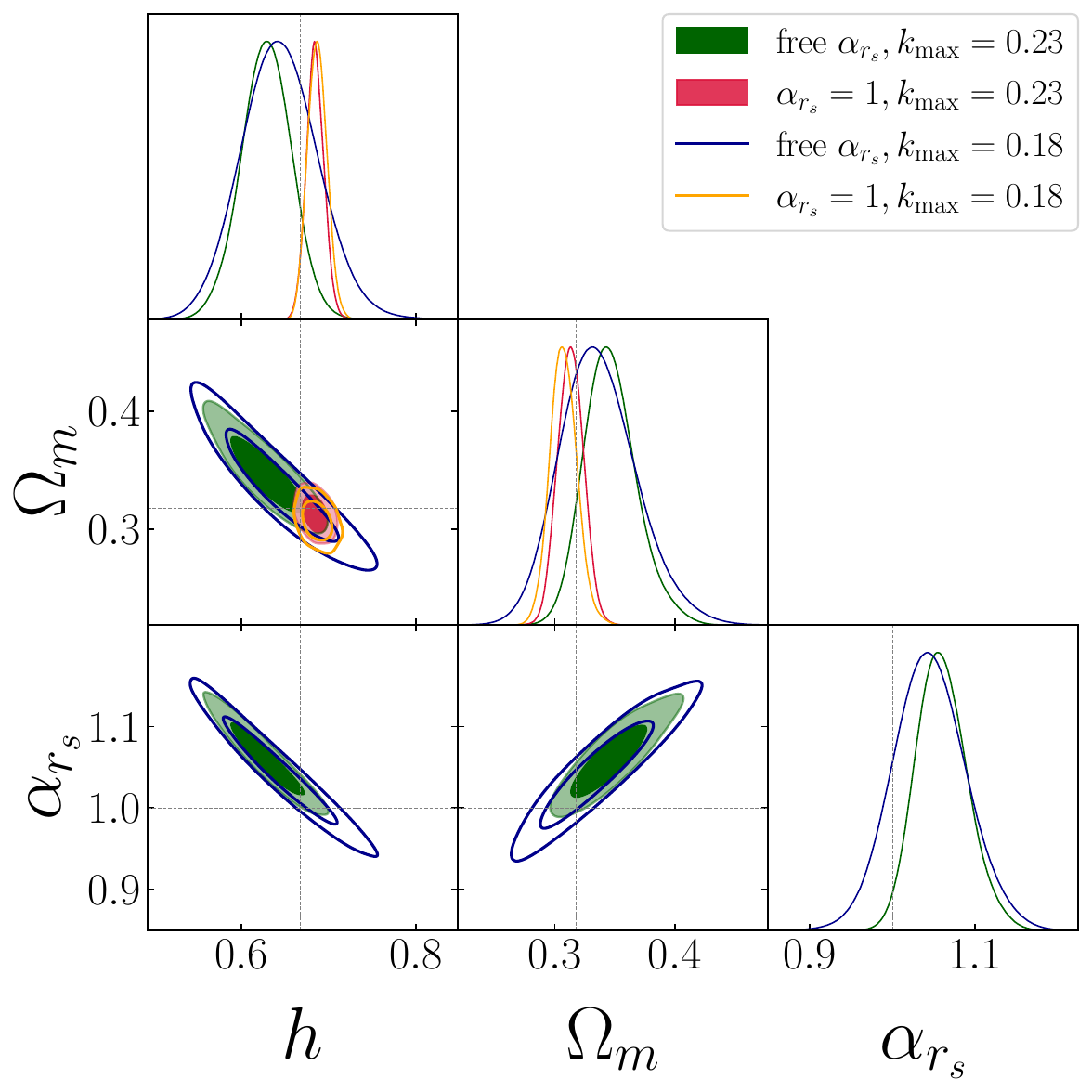}
  \caption{2D posterior distributions from the $\Lambda$CDM fit to the EFTBOSS 2+3pt mock data generated with the early dark energy cosmology (left panel) and evolving dark energy cosmology (right panel) displayed in Tab.~\ref{table:mock_cosmology}. The fiducial cosmologies are shown in dashed lines. We show the reconstructed parameters from both the $\alpha_{r_s}$-free and the $\alpha_{r_s}$-fix analyses, with the two values of $k^B_{\rm max}$ considered in this work, fixing $n_s = 0.965$ and $\omega_b = 0.02235$.}
  \label{fig:mcmc_mock_test}
\end{figure*}

It has been shown that the comparison between analyzes that marginalizes over or fix $\alpha_{r_s}$ is a useful diagnosis of the presence of physics beyond $\Lambda$CDM in the large-scale structure data. 
As an example, Ref.~\cite{Farren:2021grl} considered data from a mock Euclid experiment that were generated to contain the signal from different Early Dark Energy (EDE)~\cite{Poulin:2018cxd} models that provide large $H_0$. 
Therein, it is demonstrated that for those mock data, when analyzed under $\Lambda$CDM (\textit{i.e.}, ignoring the presence of new physics), a large shift between the inferred value of $H_0$ can appear between the $\alpha_{r_s}$-fix and $\alpha_{r_s}$-free analyses. 
Such a shift is a signature of a bias induced by using the wrong model to analyze the data.

In order to assess whether the shift we have found in the EFTBOSS likelihood when $k_{\rm max}^B = 0.23\,h\,{\rm Mpc}^{-1}$ could also entail such a signature, we perform a set of mock analyses for an EDE cosmology and a dynamical dark energy cosmology~\cite{Cai:2025mas,Cai:2009zp,Feng:2004ad,Xia:2007km,Zhao:2012aw,Guo:2004fq,Zhao:2005vj,Xia:2005ge} with parameters specified in Tab.~\ref{table:mock_cosmology}, chosen to be representative examples from recent analyses. 
Our results are shown in Fig.~\ref{fig:mcmc_mock_test} for the two different scale cuts we consider in this work, for EDE on the left panel and for $w_0w_a$ on the right panel. We show in green the $\alpha_{r_s}$-free analyses and in red the 
$\alpha_{r_s}$-fix analyses.
There are two particularly interesting results coming out of this exercise. 

First of all, for the EDE model, we find no differences between the values of $h$ and $\Omega_m$ reconstructed in the $\alpha_{r_s}$-fix or $\alpha_{r_s}$-free analyses. Moreover, the reconstructed values are  compatible with the $\Lambda$CDM expectation, despite the mock being generated at values significantly different from $\Lambda$CDM.
This illustrates the limitation of performing analyses with data of this precision. One cannot conclude that models affecting the sound horizon and leading to large $h$ values are excluded solely based on a $\Lambda$CDM analysis: one {\it expects} to reconstruct low $h$ in that case (see also Ref.~\cite{Smith:2022iax} for similar results).  This conclusion is valid regardless of the value of the scale cut.

Second, for the $w_0w_a$ model, when including information up to $k_{\rm max}^B = 0.23\,h\,{\rm Mpc}^{-1}$, we see a difference of $\sim 1.9\sigma$ between the reconstructed $h$ values, similar in amplitude to what is found with real data ($\sim 2\sigma)$, with the $\alpha_{r_s}$-free analysis being biased low and the $\alpha_{r_s}$-fix analysis being (slightly) biased high. The difference then reduces to $1\sigma$ when the scale cut is restricted to $k_{\rm max}^B = 0.18\,h\,{\rm Mpc}^{-1}$  Mpc$^{-1}$.
{This result is consistent with Ref.~\cite{Lu:2025sjg} which shows a preference for dynamical dark energy when the bispectrum BOSS data above $k^B_{\rm max}$ are included.}
{This indicates that the behavior we see with real data is consistent with the preference for dynamical dark energy observed in the data}~\cite{Lu:2025gki,Lu:2025sjg}.

\section{Conclusion}\label{sec:conclusion}
In this work, we have carried out a comprehensive sound horizon independent multiprobe cosmological analysis using large-scale structure data from BOSS, DESI, and DES. Our main conclusions can be summarized as follows:
\begin{itemize}
    \item {Individual likelihoods exhibit mild internal tensions up to $2.6\sigma$ (between EFTBOSS and EFTDES) when marginalizing over the sound horizon, which is further reduced when assuming evolving dark energy instead of $\Lambda$CDM.
    The constraints on $h$ and $\sigma_8$ are dominated by EFTBOSS, while the constraint on $\Omega_m$ is dominated by EFTDES.}
    \item Our EFTBOSS analysis indicates a slight deviation of $\alpha_{r_s} > 1$ at $1.8\sigma$, accompanied by a $1.8\sigma$ ($1.4\sigma$) shift in $h$ ($\Omega_m$) compared with the analysis that does not marginalize over the sound horizon. 
    This deviation is due to the small scales of the bispectrum $k^B_{\rm max} > 0.18  \, h \, {\rm Mpc}^{-1}$. We further note that including the bispectrum improves the constraint  in the $\{h, \, \Omega_m \}$ plane by a factor of $\sim 2$.
    \item The combination of EFTBOSS, DESI$C_\ell$, and EFTDES provides a sufficient statistical power to constrain $h$, $\Omega_m$, and $\sigma_8$ at the $3-4\%$ level without relying on sound horizon information. This represents a significant improvement compared with previous sound horizon-free analyses, namely an improvement by a factor of $\sim 3$ compared with Lensing + PanPlus~\cite{Baxter:2020qlr}, and an improvement by a factor of $1.5$ compared with  Lensing + PanPlus + EFTBOSS 2pt~\cite{Philcox:2022sgj,Smith:2022iax}. Our sound horizon independent determination of $H_0$ lies between the values inferred from CMB and calibrated supernovae measurements, exhibiting a $1.2\sigma$ tension with \textit{Planck} and a $2.1\sigma$ tension with SH0ES (see Fig.~\ref{fig:h_ladder}).
    \item We finally perform a $\Lambda$CDM fit to EFTBOSS mock data built from an early dark energy (EDE) cosmology and a dynamical dark energy (DDE) cosmology.
    While for the EDE case, we find no difference in the $\Lambda$CDM reconstructed parameters between the sound horizon-free analysis and the full analysis, we find a difference of $\sim 1.9 \sigma$ between the reconstructed $h$ values in the DDE case.
    This indicates that models affecting the sound horizon and leading to large value of $h$ cannot be excluded solely based on a $\Lambda$CDM analysis, while we can observe a (slight) inconsistency within the $\Lambda$CDM model when fitting to a DDE cosmology. {This suggests that the behavior observed with real data is consistent with the recent hint of DDE found when analyzing those data and including information from the sound horizon \cite{Lu:2025sjg}.}
\end{itemize}

The methodology developed in this work is directly applicable to upcoming surveys such as Euclid~\cite{EUCLID:2011zbd} and LSST~\cite{LSST:2008ijt}. With their larger sky coverage and improved statistical precision, these surveys are expected to deliver significantly tighter sound horizon independent constraints and may help clarify the origin of current cosmological tensions.

\begin{acknowledgments}
We thank Pierre Zhang for valuable suggestions and discussions.
ZL acknowledges the hospitality of Tsinghua University and the National Astronomical Observatory of China during the completion of this work. 
This work was supported in part by the National Key R\&D Program of China (2021YFC2203100), by the National Natural Science Foundation of China (12433002, 12261131497), by CAS young interdisciplinary innovation team (JCTD-2022-20), and by 111 Project (B23042). VP is supported by funding from the European Research Council (ERC) under the European Union's HORIZON-ERC-2022 (grant agreement no. 101076865). TS and VP acknowledges the European Union’s Horizon Europe research and innovation programme under the Marie Skłodowska-Curie Staff Exchange grant agreement No 101086085 – ASYMMETRY.
This work was partially supported by the computational resources from the  LUPM’s cloud computing infrastructure founded by Ocevu labex and France-Grilles.
\end{acknowledgments}

\appendix

\section{Assessing the robustness of our sound horizon-free analysis}\label{sec:test_prior_and_code}

In this appendix, we investigate the impact of (i) the choice of priors on the cosmological parameters $\omega_b$ and $n_s$, and (ii) the EFT likelihood used for the EFTBOSS analysis.

\begin{figure*}[!htbp]
    \centering
    \includegraphics[width=0.45\linewidth]{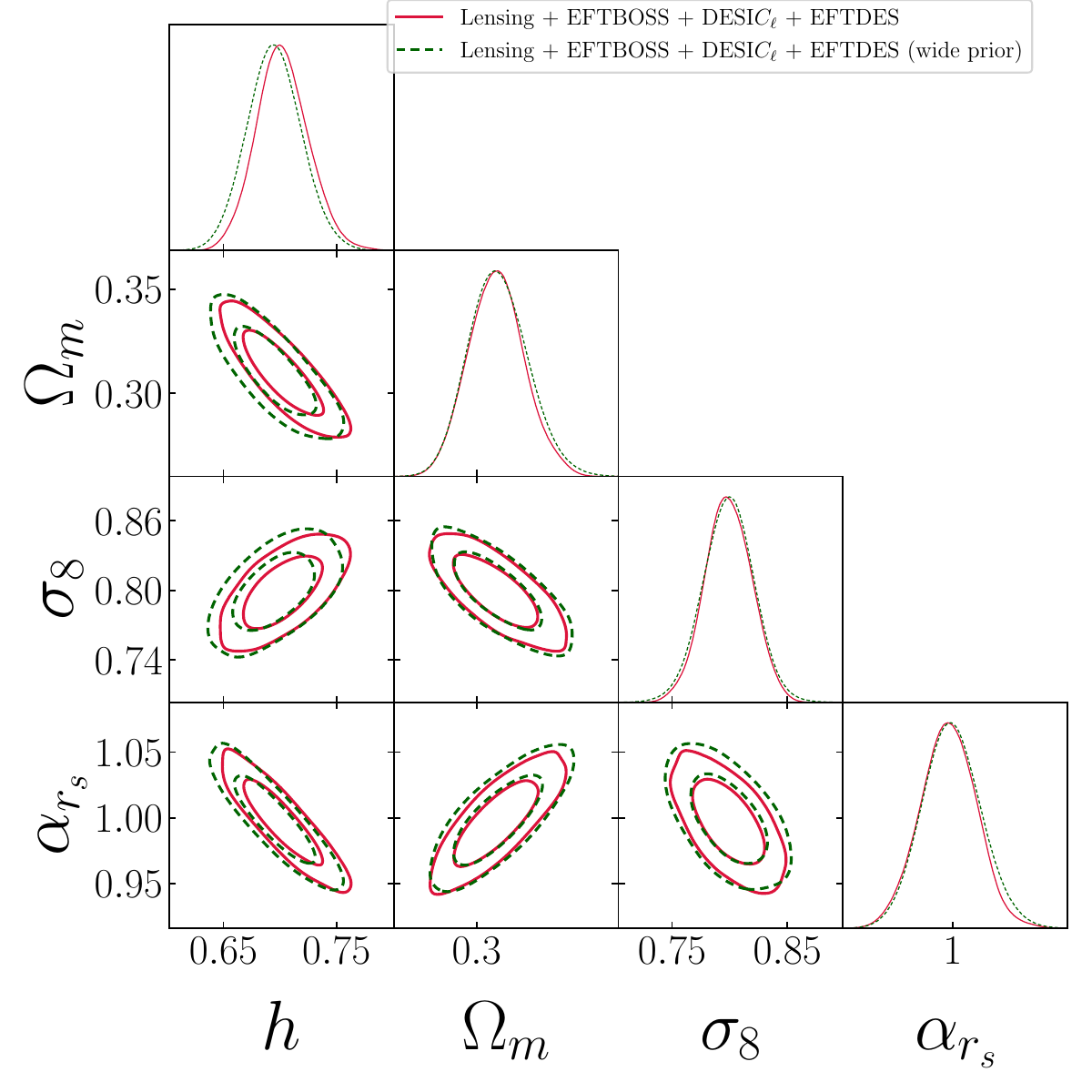}
    \includegraphics[width=0.45\linewidth]{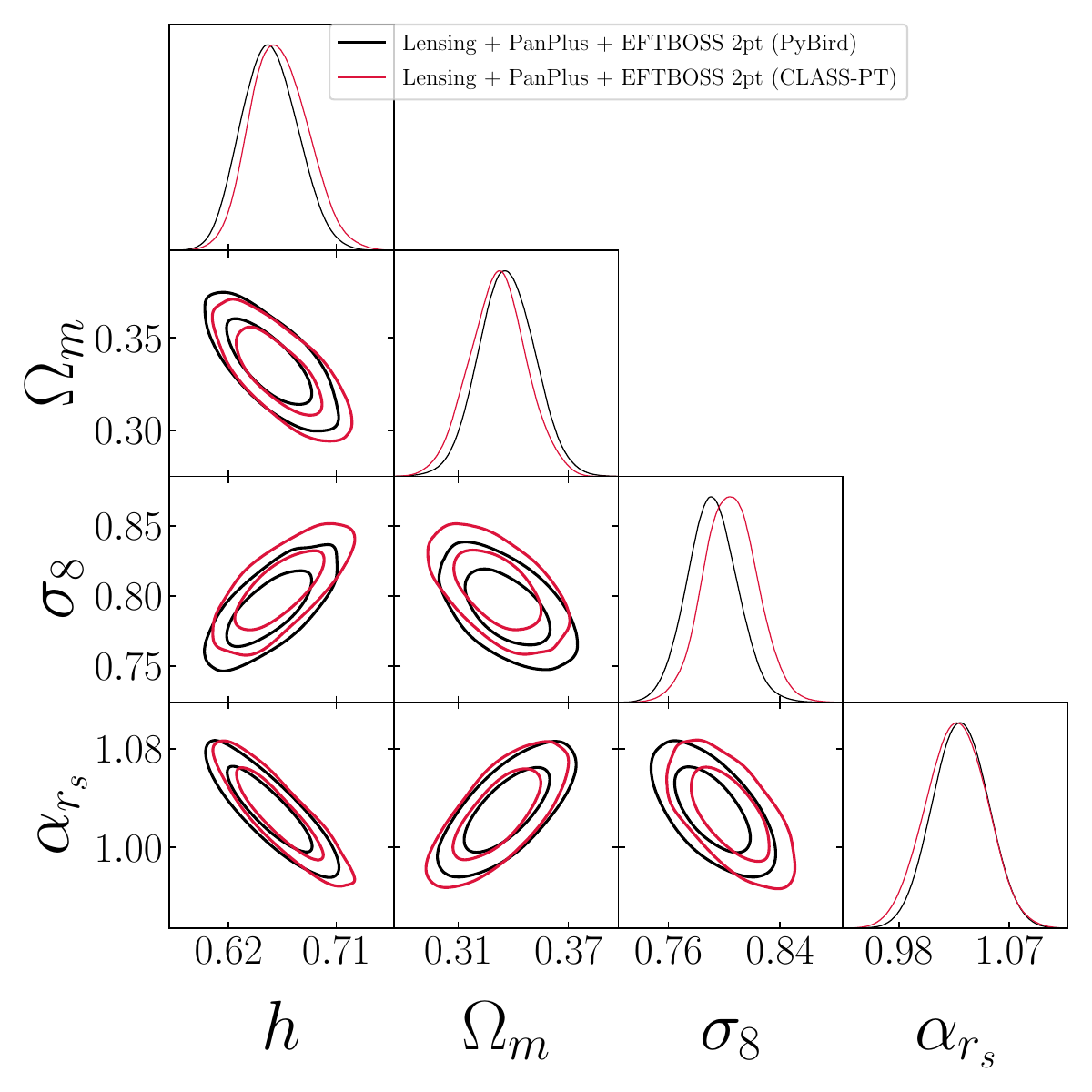}
    \caption{\textit{Left:} 2D posterior distributions from our baseline analysis, using the $\omega_b$ and $n_s$ priors specified in Sec.~\ref{sec:data} [$\omega_b\sim\mathcal N(0.02268, 0.00038)$ and $n_s\sim \mathcal N(0.96,0.02)$], and the same analysis considering different (and wider) priors on these parameters [$\omega_b\sim\mathcal N(0.02218,0.00055)$ and $n_s\sim\mathcal N(0.9649,0.042)$]. \textit{Right:} 2D posterior distributions from the galaxy power spectrum $\alpha_{r_s}$-free EFTBOSS analysis (combined with Lensing + PanPlus) obtained either with \texttt{PyBird} or \texttt{CLASS-PT}.}
    \label{fig:mcmc_add_sn_test_prior}
\end{figure*}

First, we test whether changing the priors from $\omega_b\sim\mathcal N(0.02268, 0.00038)$ and $n_s\sim \mathcal N(0.96,0.02)$ included in our baseline analysis to $\omega_b\sim\mathcal N(0.02218,0.00055)$ and $n_s\sim\mathcal N(0.9649,0.042)$, following Ref.~\cite{Zaborowski:2024car}, can affect our cosmological constraints.  The left panel of Fig.~\ref{fig:mcmc_add_sn_test_prior} shows that the resulting changes in the posterior distributions are negligible, namely $\lesssim 0.3 \sigma$. This indicates that our results are largely insensitive to the $\omega_b$ prior, confirming that our analysis remains effectively independent of information from the sound horizon.

Second, we cross-check our sound horizon-free EFTBOSS analysis, obtained with \texttt{Pybird}~\cite{DAmico:2020kxu} (using the West-Coast basis~\cite{DAmico:2020tty,DAmico:2019fhj}), against an independent pipeline, \texttt{CLASS-PT}~\cite{Chudaykin:2020aoj,Philcox:2021kcw} (using the East-Coast basis~\cite{Mirbabayi:2014zca,Fujita:2020xtd}). While previous works, such as ~\cite{Nishimichi:2020tvu,Simon:2022lde,Holm:2023laa}, compared these two codes when $\alpha_{r_s}$ is fixed, we compare here the consistency between \texttt{Pybird} and \texttt{CLASS-PT} for an $\alpha_{r_s}$-free analysis using the galaxy power spectrum (2pt) in the right panel of Fig.~\ref{fig:mcmc_add_sn_test_prior}.
Despite differences in EFT basis conventions and numerical implementations between the two codes (see details in Ref.~\cite{Simon:2022lde}) leading to results that differ at the $1 \sigma$ level on the $\Lambda$CDM parameters (in particular on $A_s$, $\sigma_8$ and $\Omega_m$) for an $\alpha_{r_s}$-fix analysis~\cite{Simon:2022lde,Holm:2023laa,Carrilho:2022mon,Gsponer:2023wpm}, we find here an excellent agreement at $\lesssim 0.5 \sigma$ for the $\alpha_{r_s}$-free analysis.

{
\section{Complementary Results}

{ In Fig.~\ref{fig:bestfit_vs_data}, we display the bestfit predictions from \texttt{PyBird} for the power spectrum and bispectrum of the BOSS data, as well as the bestfit predictions from \texttt{Swift$C_\ell$} for the DESI Legacy Survey galaxy-galaxy angular power spectra.}

\begin{figure*}[!htbp]
    \centering
    \includegraphics[width=0.45\linewidth]{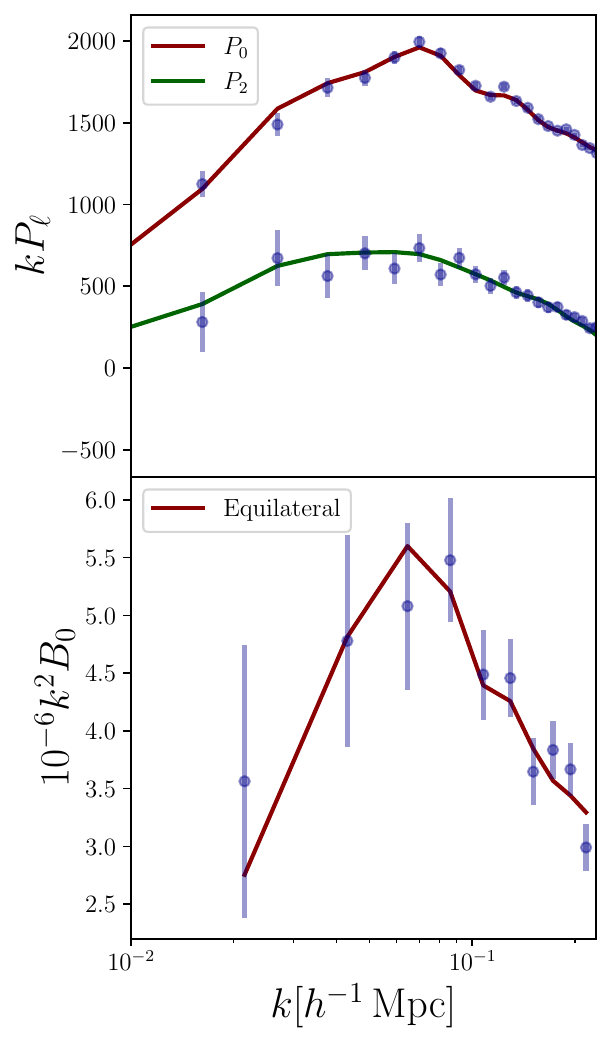}
    \includegraphics[width=0.45\linewidth]{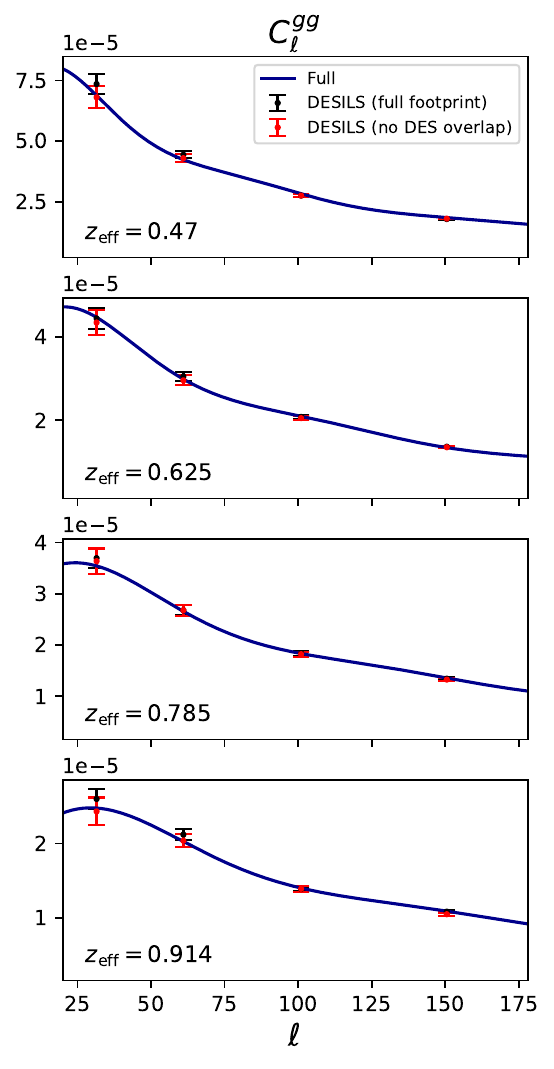}
    \caption{{\textit{Left:} Bestfit predictions for the power spectrum and bispectrum of the BOSS data. For simplicity, only the CMASS NGC sample is shown, and only equilateral triangles ($k_1 = k_2 = k_3$) are plotted. \textit{Right:} Bestfit predictions of the galaxy-galaxy angular power spectra fitted to the DESI Legacy Survey (DESILS) excluding the DES overlap (in red). For comparison, the full-sky measurement is shown in black.} The bestfit values are from the data combination of "Lensing + DESI$C_\ell$ + EFTDES + EFTBOSS + PanPlus + DESI BAO" for $\Lambda$CDM+$\alpha_{r_s}$ in table.~\ref{tab:cosmo_results}.
    }
    \label{fig:bestfit_vs_data}
\end{figure*}

}

\bibliographystyle{apsrev4-1}
\bibliography{bib} 
\end{document}